\documentclass[twocolumn]{aastex62}
\usepackage{natbib}
\defcitealias{2015MNRAS.451.2123T}{TLP15}

\received{February 20, 2020}
\accepted{June 11, 2020}
\submitjournal{ApJ}

\shorttitle{Radio emission from ultra-stripped SNe}
\shortauthors{Matsuoka et al.}

\begin{document}

\title{Radio Emission from Ultra-stripped Supernovae as Diagnostics for Properties of the Remnant Double Neutron Star Binaries}

\correspondingauthor{Tomoki Matsuoka}
\email{t.matsuoka@kusastro.kyoto-u.ac.jp}

\author{Tomoki Matsuoka}
\affil{Department of Astronomy, Kyoto University, \\
Kitashirakawa-Oiwake-cho, Sakyo-ku, Kyoto, 606-8502, Japan}

\author{Keiichi Maeda}
\affiliation{Department of Astronomy, Kyoto University, \\
Kitashirakawa-Oiwake-cho, Sakyo-ku, Kyoto, 606-8502, Japan}

\begin{abstract}
An ultra-stripped supernova (SN) is an explosion of a helium or C+O star whose outer envelope has been stripped away by a companion neutron star. 
A double neutron star (DNS) binary is believed to be left after the explosion, which {will} emit the gravitational wave later at the coalescence. 
Recent detections of a few candidates {for the} ultra-stripped SN have constrained the properties of the explosion and the progenitor, but little information is given as {to} whether the remnant DNS binary will merge within the cosmic age. A large fraction of the material stripped away from the helium star through the binary interaction is expected to escape from the system and form circumstellar material (CSM). The CSM {should} be traced by radio emission induced by the collision {with} the SN ejecta. {Based on the stellar evolution models previously developed,} we calculate the expected radio luminosities from ultra-stripped SNe. We find that {high} radio luminosity {at its maximum} can be {an indicator} of {small} separation of a DNS binary {leading to {its} merger within the cosmic age.} Our results can be used to optimize the {strategy {for} the radio follow-up observations such as observational epochs and frequencies.} 
\end{abstract}


\section{Introduction} \label{sec:intro}
The existence of double neutron star (DNS) binaries has been confirmed by the detection of the gravitational wave and the electromagnetic counterpart from a DNS merger \citep[e.g.,][]{Monitor:2017mdv, TheLIGOScientific:2017qsa, GBM:2017lvd, 2017ApJ...848L..17C, 2017PASJ...69..102T}, as well as by the direct observations of radio pulsars \citep[e.g.,][]{2003Natur.426..531B}. In the formation process of the DNS binary, the system must experience the core-collapse supernova (SN) twice, which is an explosion of a massive star at the endpoint of the stellar evolution. Hence, studies on DNS binaries provide us with the information on the stellar evolution of massive stars involved in a binary system \citep[for a review, see][]{1991PhR...203....1B}.

One of the leading models for the {formation of DNS binaries} is the ultra-stripped SN scenario \citep[][]{2017ApJ...846..170T}. In a close binary consisting of two massive stars, the secondary star loses its hydrogen envelope through the common envelope interaction with the companion NS after the first SN explosion. Subsequently, {even} the helium layer of the secondary star is {fully or partly} stripped away by the Roche lobe overflow (RLO). The secondary star then explodes as an ultra-stripped SN. This evolution scenario {leads to small} ejecta mass in the second SN, which is crucial for the binary to survive as a DNS binary system. {The characteristics of the progenitor, nucleosynthesis, and expected observational properties {of the ultra-stripped SNe in the optical wavelength} have been theoretically investigated} \citep{2013ApJ...778L..23T, 2015MNRAS.451.2123T, 2017MNRAS.466.2085M}. 

Thanks to the development of high-cadence transient surveys and {fast-turnaround} follow-up observations, a few candidates {for} ultra-stripped SNe have been discovered \citep[e.g., iPTF 14gqr,][]{2018Sci...362..201D}. {The timescale of the optical light curve of iPTF 14gqr is $\sim 5$ days, whereas those of typical Type Ib/Ic SNe are $10 \sim 20$ days \citep{2016MNRAS.457..328L}. This implies that the ejecta mass of iPTF 14gqr is small ($\sim 0.1 M_\odot$).} The maximum-light spectrum is reminiscent of {those of} Type Ic SN{e}, indicating {that the progenitor is a C+O star.} These observational features agree with the prediction for the ultra-stripped SN \citep{2017MNRAS.466.2085M}. {Furthermore, a few other candidates have been suggested from the viewpoints of their spectra and rapidly decaying evolutions, including SN 2005ek \citep{2013ApJ...774...58D, 2013ApJ...778L..23T} and SN 2010X \citep{2010ApJ...723L..98K}\footnote{See \citet{2017MNRAS.466.2085M} and \citet{2020arXiv200502992N} for the other candidates.}. We note that for SN 2010X a progenitor model originated from a white dwarf has also been suggested}.

However, the remnant DNS binaries do not {necessarily} have sufficiently small separations to merge within the cosmic age \citep[][see also Section \ref{sec:progenitors}]{2015MNRAS.451.2123T}. In fact, a DNS binary with a long orbital period ($\sim 45$ days, corresponding to the separation of $\sim 0.4$ AU) has been {discovered} by the radio pulsar observation \citep{2015ApJ...805..156S}. Optical properties of ultra-stripped SNe are {sensitive to the ejecta mass \citep{2017MNRAS.466.2085M}, but not to the separation of the remnant DNS binary. Ultra-stripped SNe with the small ejecta mass ($\lesssim 0.2M_\odot$) can {indeed} originate in a wide range of the binary separation (see Section \ref{sec:progenitors} for details).} Therefore, independent diagnostics {for} the remnant DNS binary separation after an ultra-stripped SN event will be important.

A key ingredient in the ultra-stripped SN scenario is the RLO mass transfer from the progenitor to the companion NS. A large fraction of the material is expected to be ejected from the system and form circumstellar material (CSM), which will lead to intense radio emission following an SN explosion \citep[e.g.,][]{1982ApJ...259..302C}. In this paper, we investigate properties of radio emission from the ultra-stripped SN-CSM interaction. {We suggest that the strong radio emission can be a tracer of an ultra-stripped SN which leaves a remnant DNS binary with sufficiently small separation to merge within the cosmic age.}

This paper is organized as follows. In Section \ref{sec:progenitors}, we review the characteristics of the ultra-stripped SN progenitors in the stellar evolution models proposed by \cite{2015MNRAS.451.2123T}. 
{We thereby find a trend that the mass-transfer rate is larger for the small binary separation, especially for the models with small ejecta mass ($<0.2M_\odot$).} In Section \ref{sec:method}, we describe the method for calculating the radio emission. The results are shown in Section \ref{sec:results}. We discuss the observational prospects, together with some limitations in the {present} models, in Section \ref{sec:discussions}. The paper is closed in Section \ref{sec:summary} with a summary of our findings.

\section{Properties of the ultra-stripped SN progenitors and mass-transfer rates} \label{sec:progenitors}
\begin{figure*}[ht!]
\plotone{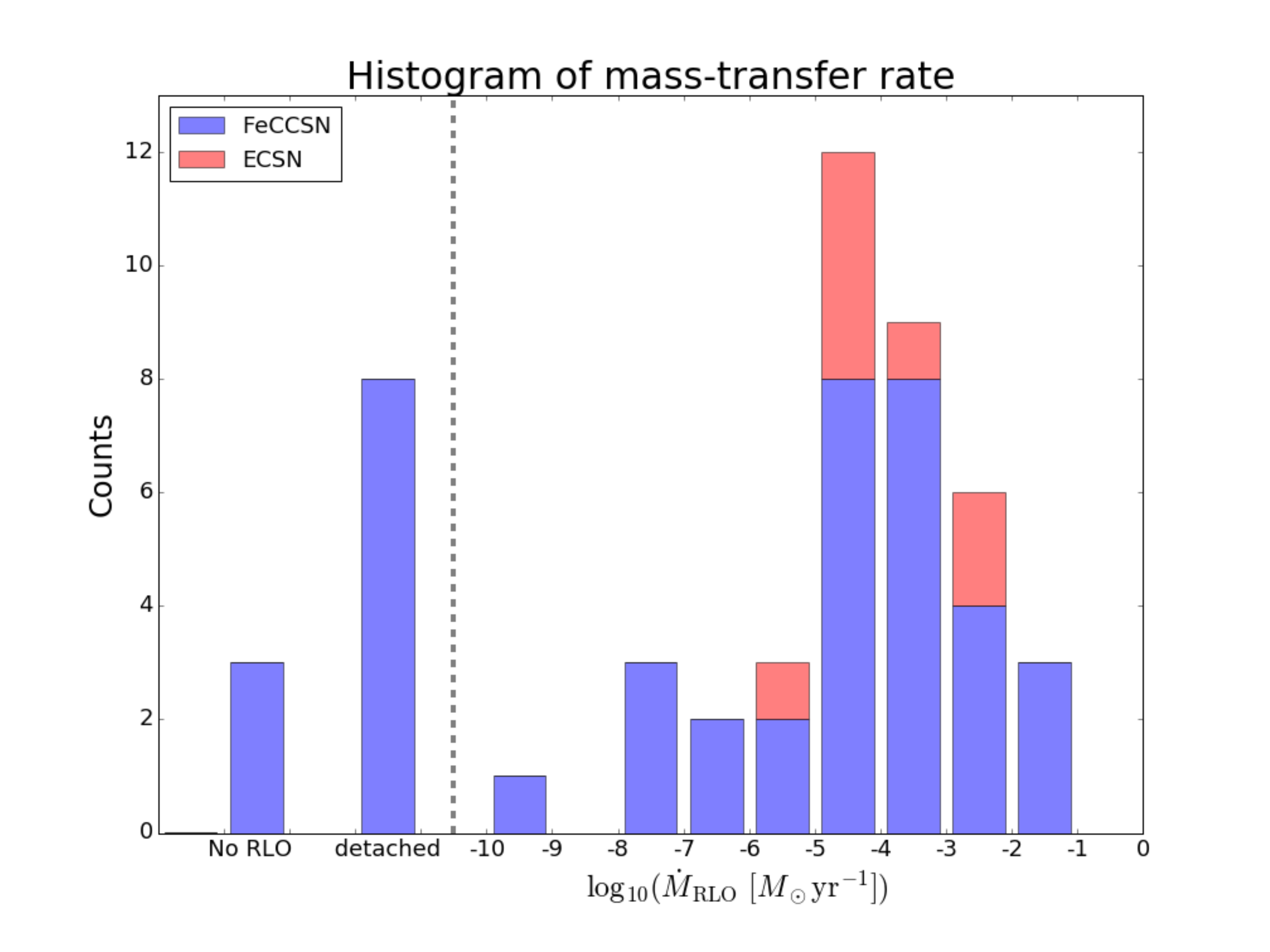}
\caption{The histogram of the final mass-transfer rates {in the models presented} by \citetalias{2015MNRAS.451.2123T}. The difference {in the color} shows the type of SNe; FeCCSN (blue) or ECSN (red). The models in which {either} the binary is detached or no RLO initiates are {separately shown} in the left side of the histogram.}
\label{fig:Histogram}
\end{figure*}

\citet[hereafter \citetalias{2015MNRAS.451.2123T}]{2015MNRAS.451.2123T} presented a series of the evolution models for a helium star with the helium envelope stripped away by a companion NS, providing a table for the final RLO mass-transfer rates and the fates of the {helium stars}. Figure \ref{fig:Histogram} shows a histogram of the final mass-transfer rates ($\dot{M}_{\rm RLO}$) reproduced from Table 1 of \citetalias{2015MNRAS.451.2123T}. For the models in which the final fate is either the iron core-collapse SN (FeCCSN) or {the} electron capture SN (ECSN), we observe that the average mass-transfer rate ($\dot{M}_{\rm RLO} \sim 10^{-4} \ M_\odot\mbox{yr}^{-1}$) is higher than the canonical mass-loss rate expected for helium stars {\citep[$\dot{M}_{\rm wind} \sim 10^{-5} \ M_\odot\mbox{yr}^{-1}$ ,][]{2006ApJ...651..381C, 2015ApJ...806..106A}}. There are some models {in which} {either} the binary is detached or no RLO initiates. In these models the CSM {around the progenitor} will be dominated by the stellar wind. Note that the models, in which {either} a white dwarf is left or {the common envelope interaction is realized,} are not included in Figure \ref{fig:Histogram}, {because} {the secondary stars in} these models would not explode as SNe. 

\begin{figure*}[]
\plotone{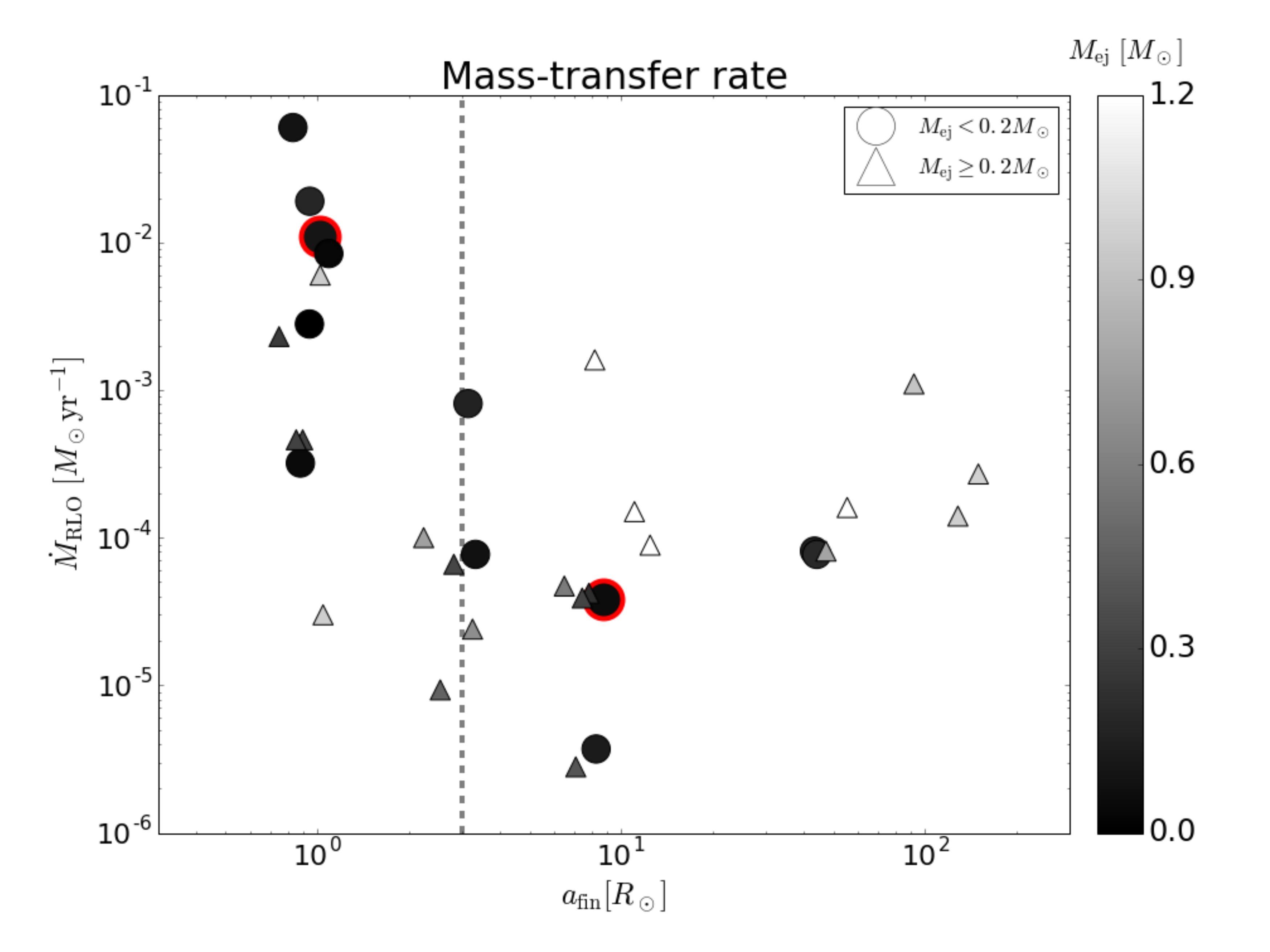}
\caption{Distribution of the mass-transfer rate {as a function of} the final separation. Only the models {in which the secondary stars} explode as SNe are plotted. {Two different symbols are used depending on} the difference in the ejected mass; $M_{\rm ej} < 0.2\ M_\odot$ (circles) or $M_{\rm ej} \geq 0.2\ M_\odot$ (triangles). The models plotted by the {circles {outlined in red}} are examined in details (see Section \ref{sec:LCs}). For the models {on} the {left side of the black dashed line,} the remnant DNS binary is expected to merge within the cosmic age.}
\label{fig:Distribution of Mdot}
\end{figure*}

Figure \ref{fig:Distribution of Mdot} shows the distribution of the final mass-transfer rate ($\dot{M}_{\rm RLO}$) {as a function of} the final separation calculated by the Keplerian law ($a_{\rm fin}$), reproduced from Table 1 of \citetalias{2015MNRAS.451.2123T}. 
{The ejecta mass in each model is estimated as follows:}

\begin{eqnarray}
M_{\rm ej} = M_{\ast,{\rm f}} - M_{\rm NS},
\end{eqnarray}
{where} $M_{\ast,{\rm f}}$ and $M_{\rm NS} (= 1.4 \ M_\odot)$ are the final mass of the helium star computed by \citetalias{2015MNRAS.451.2123T} and the mass of the newly born NS, respectively. 

{In {Figure \ref{fig:Distribution of Mdot}} we find two characteristics on the mass-transfer rate and the final binary separation. First, there are some models located at $(a_{\rm fin}, \dot{M}_{\rm RLO}) \sim (1R_\odot, 10^{-2}M_\odot{\rm yr}^{-1})$. This separation satisfies the condition that the remnant DNS binary will merge within the cosmic age ($a_{\rm fin} \lesssim 3.3\ R_\odot$). The mass-transfer rate is at least by an order of magnitude larger than those of the models with the larger separations. In fact, one of the models with the parameter set $(a_{\rm fin}, \dot{M}_{\rm RLO}) \sim (1R_\odot, 10^{-2}M_\odot{\rm yr}^{-1}$), shown by the circle {outlined in red} in the top left of Figure \ref{fig:Distribution of Mdot}, has been adopted to explain the optical properties of an ultra-stripped SN candidate SN 2005ek \citep{2013ApJ...778L..23T}. Hence, we conclude that 
if the very high mass-transfer rate ($\gtrsim 10^{-2} M_\odot{\rm yr}^{-1}$) is derived for a progenitor of an ultra-stripped SN candidate through the property of the CSM, this will infer that the binary is sufficiently close for the remnant DNS binary to make a coalescence within the cosmic age.

Second, for the models with $M_{\rm RLO} \sim 10^{-4} M_\odot{\rm yr}^{-1}$, a range of final separations ($a_{\rm fin}$) could be associated. However, if we focus only on the models with the small ejecta mass ($M_{\rm ej} < 0.2 M_\odot$), there is a tendency for $\dot{M}_{\rm RLO}$ to decrease as $a_{\rm fin}$ increases. Therefore, in case the small ejecta mass is derived through the optical properties, the mass-loss rate could be used as a rough tracer of the DNS binary separation.}

In summary, we see the following trend on the mass-transfer rate. The ultra-stripped SN progenitor tends to have a higher mass-transfer rate than the typical mass-loss rate of a helium star. {Especially, for the model with which an extraordinarily high mass-transfer rate ($\dot{M}_{\rm RLO} \gtrsim 10^{-3} M_\odot{\rm yr}^{-1}$) is associated, the binary separation is so small that the remnant DNS binary will make a coalescence within the cosmic age.} 

\section{Models and Method}\label{sec:method}
After the shock breakout, the collisionless shock is formed. It is a site for non-thermal particle acceleration and magnetic field amplification, followed by multi-wavelength emissions including synchrotron radio emission. In this section we describe the method to model the synchrotron emission, which basically follows the previous studies \citep[e.g.,][]{1982ApJ...259..302C, 1998ApJ...499..810C, 1998ApJ...509..861F, 2006ApJ...651..381C, 2017hsn..book..875C, 2012ApJ...758...81M, 2013ApJ...762...14M}.

\subsection{Models}\label{sec:models}
The evolution of the shock velocity is determined by the natures of the SN ejecta and the CSM. The outer ejecta structure can be described by a power-law, $\rho \propto t^{-3} v^{-n}$, where $v=r/t$ is the velocity coordinate. In case that the outer envelope of the progenitor is radiative, the power-law index $n = 11.73$ matches to the density structure of the outermost layer of the SN ejecta \citep{1999ApJ...510..379M}. The ultra-stripped SN progenitors {likely have} a radiation-dominant envelope \citep[e.g.,][]{2015MNRAS.454.3073S}, and we employ the same index.

{For the ejecta mass exceeding $\sim 0.2 M_\odot$, the evolution of the optical light curve becomes too slow to be consistent with the rapid evolutions seen in the previously observed candidates for the ultra-stripped SNe \citep{2017MNRAS.466.2085M, 2018Sci...362..201D}.} Thus, in this study, we impose a threshold on the ejecta mass, and focus on the models {in} which the ejecta mass is smaller than $0.2 M_\odot$ (the circles in Figure \ref{fig:Distribution of Mdot}).

With the ejecta properties given, the synchrotron emission can be computed once the properties of the CSM are specified. A fraction of the {gas transferred from} the helium star to the NS is expected to escape from the binary system. This material will be distributed around the binary as the CSM. In this study, we parametrize the CSM density distribution $\rho_{\rm CSM}(r)$ as follows:
\begin{eqnarray}
\rho_{\rm CSM}(r) = \frac{\dot{M}_{\rm CSM}}{4\pi u_{\rm w}}\frac{1}{r^2}, \mbox{ and } \dot{M}_{\rm CSM} = f_{\dot{M}} \dot{M}_{\rm RLO},\label{eq:CSM}
\end{eqnarray}
where $u_{\rm w}$ and $f_{\dot{M}}$ are the mass-loss velocity and the fraction of the gas {escaping} from the system, respectively.
 
The value of $u_{\rm w}$ involves a large uncertainty. We adopt the typical escape velocity from a helium star, $u_{\rm w} = 10^8 \mbox{ cm s}^{-1}$ \citep{2000A&A...360..227N}, {because this value is larger than the binary orbital velocity ($\sim 2\pi a_{\rm fin} P_{\rm fin}^{-1} \sim 10^7$ cm s$^{-1}$, where $P_{\rm fin}$ is the final orbital period). However, we note that the velocity of the outflow from the NS might be even larger than the typical escape velocity from a helium star} \citep{2016ApJ...822L..18M}. 

$f_{\dot{M}}$ is also an important parameter which determines the efficiency of {the formation of} the CSM around the progenitor. {It is expected that $f_{\dot{M}}$ is large ($f_{\dot{M}}\sim 1$) for the following reasons. First, the mass-transfer rate here is a few orders of magnitude higher than the Eddington accretion rate onto a NS ($\dot{M}_{\rm Edd, NS} \sim 10^{-8} \ M_\odot\mbox{yr}^{-1}$). Therefore, most of the materials cannot accrete onto the NS, and will escape from the binary system. This is also required in order for the NS to avoid a collapse to a black hole. Second, the NS should be spinning up rapidly under the ultra-stripped SN scenario, and it can no longer receive the angular momentum from the accreting gas.} We thus consider $f_{\dot{M}} = 0.99$ as our fiducial model. We also examine $f_{\dot{M}} = 0.10$ to investigate the dependence of the radio emission on this parameter. Finally, we remark that any asphericity of the CSM is not considered in this work, although the gas {escaping from the system will not} necessarily be distributed spherically. 

\subsection{Shock evolution}\label{sec:hydrodynamics}
Assuming that the shocked region is geometrically thin, {the velocity ($V_{\rm sh}$) and the radius ($R_{\rm sh}$)} of the shocked shell can be derived analytically as follows \citep{1982ApJ...258..790C,1982ApJ...259..302C}:
\begin{eqnarray}
V_{\rm sh} &=& {1.1}\times {10^9} 
{\displaystyle \left(
\frac{\dot{M}_{\rm CSM}}{{10^{-2}}\ M_\odot \mbox{yr}^{-1}}
\right)^{-0.10}
\left(
\frac{u_{\rm w}}{10^8 \mbox{ cm s}^{-1}}
\right)^{0.10}} \nonumber \\
&\times&
{\displaystyle \left(
\frac{E_{\rm kin}}{10^{50} \mbox{ erg}}
\right)^{0.45}
\left(
\frac{M_{\rm ej}}{0.1\ M_\odot}
\right)^{-0.35}
\left(
\frac{t}{10\mbox{ days}}
\right)^{-0.10} } \label{eq:V_sh}\nonumber \\
&&\mbox{\hspace{60mm}cm s}^{-1},\\
R_{\rm sh} &=& {8.5}\times {10^{14}}
{\displaystyle \left(
\frac{\dot{M}_{\rm CSM}}{10^{-2}\ M_\odot \mbox{yr}^{-1}}
\right)^{-0.10}
\left(
\frac{u_{\rm w}}{10^{8} \mbox{ cm s}^{-1}}
\right)^{0.10}} \nonumber \\
&\times&
{\displaystyle \left(
\frac{E_{\rm kin}}{10^{50} \mbox{ erg}}
\right)^{0.45}
\left(
\frac{M_{\rm ej}}{0.1\ M_\odot}
\right)^{-0.35}
\left(
\frac{t}{10\mbox{ days}}
\right)^{0.90}} \label{eq:R_sh}\nonumber \\
&&\mbox{\hspace{60mm}cm s}^{-1},
\end{eqnarray}
where $\dot{M}_{\rm CSM}$ and $E_{\rm kin}$ are the mass-loss rate converted from the CSM density (see above) and the kinetic energy of the ejecta, respectively. We do not take radiative cooling into account, which could {decelerate the shock velocity by roughly a ten percent if the CSM density is {high}. The radio light curves are hardly affected by this assumption \citep{2019ApJ...885...41M}.}

\subsection{Particle Acceleration and Magnetic Field Amplification}\label{sec:B_and_N}
At the collisionless shock front, charged particles such as electrons or protons become energetic by diffusive shock acceleration \citep[DSA,][]{1949PhRv...75.1169F, 1978MNRAS.182..147B, 1983RPPh...46..973D}. The motion of the charged particles is relativistic and random, followed by magnetic field amplification. We parametrize the energy density of the electrons ($u_e$) and magnetic field ($u_B$) as a fraction of the post-shocked energy density as a function of time as follows:
\begin{eqnarray}
u_e = \epsilon_e \rho_{\rm sh} V_{\rm sh}^2, \label{eq:5} \\
u_B = \frac{B^2}{8\pi} =\epsilon_B \rho_{\rm sh} V_{\rm sh}^2, \label{eq:6}
\end{eqnarray}
where $\rho_{\rm sh}$ is the post-shocked density of the CSM, which is 4 times larger than that of the pre-shocked CSM. $\epsilon_e$ and $\epsilon_B$ are the parameters which determine the efficiency of the shock acceleration and the magnetic field amplification. In this study we use the values $\epsilon_e = 0.01$ and $\epsilon_B = 0.1$, but we note that there remains a debate on {the realistic values of these parameters} \citep[e.g.,][]{2008ApJ...682L...5S, 2012ApJ...758...81M, 2015ApJ...798L..28C}.

We consider the power-law distribution of the {number density of the} accelerated electrons $N$ as a function of the Lorentz factor ($\gamma$) as follows:
\begin{eqnarray}
N(\gamma) = C \gamma^{-p}.
\end{eqnarray}
The index $p$ characterizes the hardness of the spectrum of the electron distribution. We employ $p=3$, which can explain observations of optically thin radio emissions from Type Ib/Ic SNe \citep{2006ApJ...651..381C, 2013ApJ...762...14M}. The coefficient $C$ is determined by equating the integrated energy density of the electrons with $u_e$,
\begin{eqnarray}
\int_{\gamma_{\rm min}}^{\infty} d\gamma N(\gamma) \gamma m_e c^2 = u_e \Rightarrow C = \frac{(p-2)u_e}{\gamma_{\rm min}^{2-p} m_e c^2} \label{eq:C},
\end{eqnarray}
where {$m_e, c$, and $\gamma_{\rm min}=2$ are the electron mass, the speed of light, and the minimum Lorentz factor of the accelerated electrons, respectively}.

Our treatment does not include the contribution from hadronic interactions. It is possible that the relativistic protons collide with target protons in the dense CSM, producing electrons and positrons via pion decay \citep[e.g.,][]{2016MNRAS.460...44P, 2014MNRAS.440.2528M, 2019ApJ...874...80M}. However, the previous simulation of the radio emission from infant Type II-P SNe has shown that the synchrotron emission from these secondary particles would receive strong self-absorption when the luminosity is at its maximum \citep{2019ApJ...885...41M}. Therefore, in this study we neglect the contribution from the hadronic interactions.

\subsection{Synchrotron emission}\label{sec:synchrotron}
Once the energy distribution of the electrons and the strength of the magnetic field are given, the physical quantities for the synchrotron emission can be calculated \citep{1979rpa..book.....R}. The emissivity $j_{\nu, {\rm syn}}$ and the synchrotron self-absorption (SSA) coefficient $\alpha_{\nu, {\rm SSA}}$ are described as follows:
\begin{eqnarray}
j_{\nu, {\rm syn}} &=& \frac{1}{4\pi}\int d\gamma P_{\nu, {\rm syn}}(\gamma) N(\gamma), \\
\alpha_{\nu, {\rm syn}} &=& -\frac{1}{8\pi \nu^2 m_e} \int d\gamma P_{\nu, {\rm syn}}(\gamma) \gamma^2 \frac{\partial}{\partial \gamma}
\left[
\frac{N(\gamma)}{\gamma^2}
\right].\nonumber \\
\end{eqnarray}
$P_{\nu, {\rm syn}}(\gamma)$ is the power per unit frequency emitted by one electron defined as follows:
\begin{eqnarray}
P_{\nu, {\rm syn}}(\gamma) &=& \frac{\sqrt{3}q^3B\sin\theta}{m_e c^2}F\left(\frac{\nu}{\nu_c}\right), \\
\nu_c = \frac{3 \gamma^2 q B \sin\theta}{4\pi m_e c}&,&\  F(x) = x\int_x^\infty K_{5/3}(y)dy,
\end{eqnarray}
where $q$ and $\theta$ are the elementary charge and the pitch angle of the electron ($\sin \theta = 2/3$), respectively. $K_{5/3}(y)$ is the modified Bessel function. For the power-law energy distribution of the electrons, the above integration is analytically calculated as follows:
\begin{eqnarray}
j_{\nu, {\rm syn}} &=& \frac{\sqrt{3}q^3CB\sin\theta}{4\pi m_e c^2}
\frac{2^{(p-1)/2}}{p+1}
\Gamma\left(\frac{3p+19}{12}\right) \nonumber \\
&\times&\Gamma\left(\frac{3p-1}{12}\right)
\left(\frac{\nu}{\nu_c(\gamma=1)}\right)^{-(p-1)/2} \nonumber \\
&\simeq& 1.50\times 10^{-23} 
\left(\frac{C}{1\mbox{ cm}^{-3}}\right)
\left(\frac{B\sin\theta}{1\mbox{ gauss}}\right)\nonumber \\
&\times&\left(\frac{\nu}{\nu_c(\gamma=1)}\right)^{-1}\mbox{ erg}\ {\rm s}^{-1}\ {\rm cm}^{-3}\ {\rm Hz}^{-1}\ {\rm str}^{-1},\nonumber \\
\ \\
\alpha_{\nu, {\rm syn}} &=& \frac{2\sqrt{3}\pi}{9} 2^{p/2} \frac{qC}{B\sin\theta}
\Gamma\left(\frac{3p+2}{12}\right) \nonumber \\
&\times&\Gamma\left(\frac{3p+22}{12}\right) 
\left(\frac{\nu}{\nu_c(\gamma=1)}\right)^{-(p+4)/2} \nonumber \\
&\simeq& 2.45\times 10^{-9}
\left(\frac{C}{1\mbox{ cm}^{-3}}\right) \nonumber \\
&\times&\left(\frac{B\sin\theta}{1\mbox{ gauss}}\right)^{-1}
\left(\frac{\nu}{\nu_c(\gamma=1)}\right)^{-7/2} \mbox{ cm}^{-1}.
\end{eqnarray}

If the CSM is very dense, free-free absorption (FFA) in the pre-shocked CSM region becomes important. The FFA absorption coefficient is estimated in cgs unit as follows:
\begin{eqnarray}
\alpha_{\nu,{\rm ff}} = 0.018 T_e^{-3/2} Z^2 \frac{\rho_{\rm CSM}^2}{\mu_i \mu_e m_p} \nu^{-2} g_{\rm ff} \mbox{ cm}^{-1},
\end{eqnarray}
where $T_e, Z, \mu_i, \mu_e, m_p$, and $g_{\rm ff}$ are the thermal electron temperature, the charge of the thermal ions, the molecular weights of ions and electrons, the proton mass, and the free-free gaunt factor, respectively. We use $T_e = 10^5$ K, which is conventionally used for explaining the absorption of radio emission from Type II SNe \citep[see e.g.,][]{1988A&A...192..221L, 2006ApJ...641.1029C}. We note, however, that $T_e$ involves large uncertainties and this could have effect on the radio emission in the early phase of SNe. The composition of the CSM is dominated by fully ionized helium, and thus $Z=2$ is used. For the free-free gaunt factor, we use the formalization described by \cite{1979rpa..book.....R}.

The observed radio luminosity per unit frequency ($L_\nu$) can be estimated by using the synchrotron source function ($S_{\nu,{\rm syn}}$), and the optical depths to SSA ($\tau_{\nu, {\rm syn}}$) and FFA ($\tau_{\nu, {\rm ff}}$), as follows \citep{1998ApJ...509..861F, 2017hsn..book..875C}:
\begin{eqnarray}
L_\nu &=& 4\pi R_{\rm sh}^2 \pi S_{\nu, {\rm syn}} (1-e^{-\tau_{\nu, {\rm syn}}})e^{-\tau_{\nu, {\rm ff}}}, \label{eq:Lnu}\\
&&\hspace{10mm}S_{\nu, {\rm syn}} = \frac{j_{\nu, {\rm syn}}}{\alpha_{\nu, {\rm syn}}} \nonumber \\
&=&6.12\times 10^{-15} 
\left(\frac{B\sin\theta}{1\mbox{ gauss}}\right)^2
\left(\frac{\nu}{\nu_c(\gamma=1)}\right)^{5/2} \nonumber \\
&&\mbox{\hspace{30mm} erg}\ {\rm s}^{-1}\ {\rm cm}^{-3}\ {\rm Hz}^{-1}\ {\rm str}^{-1}, \\
\tau_{\nu, {\rm syn}} &=& \int_{\rm shock} dr \alpha_{\nu, {\rm syn}} \nonumber \\ 
&\simeq& 0.61
\left(\frac{C}{1\mbox{ cm}^{-3}}\right)
\left(\frac{B\sin\theta}{1\mbox{ gauss}}\right)^{-1} \nonumber \\
&\times&
\left(\frac{V_{\rm sh}}{10^9 \mbox{cm s}^{-1}}\right) 
\left(\frac{t_{\rm cool}(\nu)}{1\ {\rm s}}\right)
\left(\frac{\nu}{\nu_c(\gamma=1)}\right)^{-7/2}, \\
\tau_{\nu, {\rm ff}} &=& \int_{R_{\rm sh}}^\infty dr \alpha_{\nu,{\rm ff}} \nonumber \\
&\simeq& 4.7\times 10^3\ 
\left(\frac{T_e}{10^5\ {\rm K}}\right)^{-3/2}
\left(\frac{\nu}{10^{10}\ {\rm Hz}}\right)^{-2}\nonumber \\
&\times&\left(\frac{\dot{M}_{\rm CSM}}{10^{-2}\ M_\odot{\rm yr}^{-1}}\right)^{2}
\left(\frac{u_{\rm w}}{10^{8}\ \mbox{cm s}^{-1}}\right)^{-2}
\left(\frac{R_{\rm sh}}{10^{15}\ {\rm cm}}\right)^{-3}. \nonumber \\
\end{eqnarray}
{$t_{\rm cool}(\nu)$ is the cooling timescale of electrons emitting a synchrotron photon at frequency $\nu$:
\begin{eqnarray}
t_{\rm cool}(\nu) = 
\left(
\frac{1}{t} + \frac{1}{t_{\rm syn}} + \frac{1}{t_{\rm coulomb}}
\right)^{-1},
\end{eqnarray}
where the timescales for three cooling processes are considered; dynamical ($t$), synchrotron ($t_{\rm syn}$), and Coulomb ($t_{\rm coulomb}$) cooling. The last two timescales are respectively estimated as follows:
\begin{eqnarray}
t_{\rm syn} &=& 3.2\times 10^7 
\left(\frac{B}{1\mbox{ gauss}}\right)^{-3/2} 
\left(\frac{\nu}{10^{10}\ {\rm Hz}}\right)^{-1/2} \mbox{ s}, \\
t_{\rm coulomb} &=& 3.0\times 10^4 
\left(\frac{B}{1\mbox{ gauss}}\right)^{-1/2} 
\left(\frac{\nu}{10^{10}\ {\rm Hz}}\right)^{1/2} \nonumber \\
&\times&\left(\frac{\dot{M}_{\rm CSM}}{10^{-2}\ M_\odot{\rm yr}^{-1}}\right)^{-1}
\left(\frac{u_{\rm w}}{10^{8}\ \mbox{cm s}^{-1}}\right)^{}
\left(\frac{R_{\rm sh}}{10^{15}\ {\rm cm}}\right)^{2} \mbox{ s}. \nonumber \\
\end{eqnarray}
Inverse Compton cooling is not taken into account, because the bolometric luminosity of ultra-stripped SNe is \added{much} fainter than {those} of typical SNe \citep[e.g.,][]{2017MNRAS.466.2085M}.}

{The calculation of the radio light curves is summarized as follows. At given time $t$, the hydrodynamical properties ($V_{\rm sh}, R_{\rm sh}, \rho_{\rm sh}$) are first specified (equations \ref{eq:V_sh} and \ref{eq:R_sh}). Then, the energy densities of the accelerated electrons and the magnetic field are calculated through the parametrizations in equations (\ref{eq:5}) and (\ref{eq:6}). The distribution of the number density of the accelerated electrons is then given by equation (\ref{eq:C}). The natures of the electrons and the magnetic field thus specified are used in the computation of the quantities relevant to synchrotron emission ($j_{\nu,{\rm syn}}, \alpha_{\nu,{\rm syn}}$, and $S_{\nu,{\rm syn}}$). As the cooling timescale of electrons $t_{\rm cool}(\nu)$ as well as optical depths to SSA and FFA are also determined, the radio luminosity per unit frequency can be calculated by equation (\ref{eq:Lnu}).}

\section{Results} \label{sec:results}
\subsection{Light curves}\label{sec:LCs}
\begin{figure*}[ht!]
\gridline{
\fig{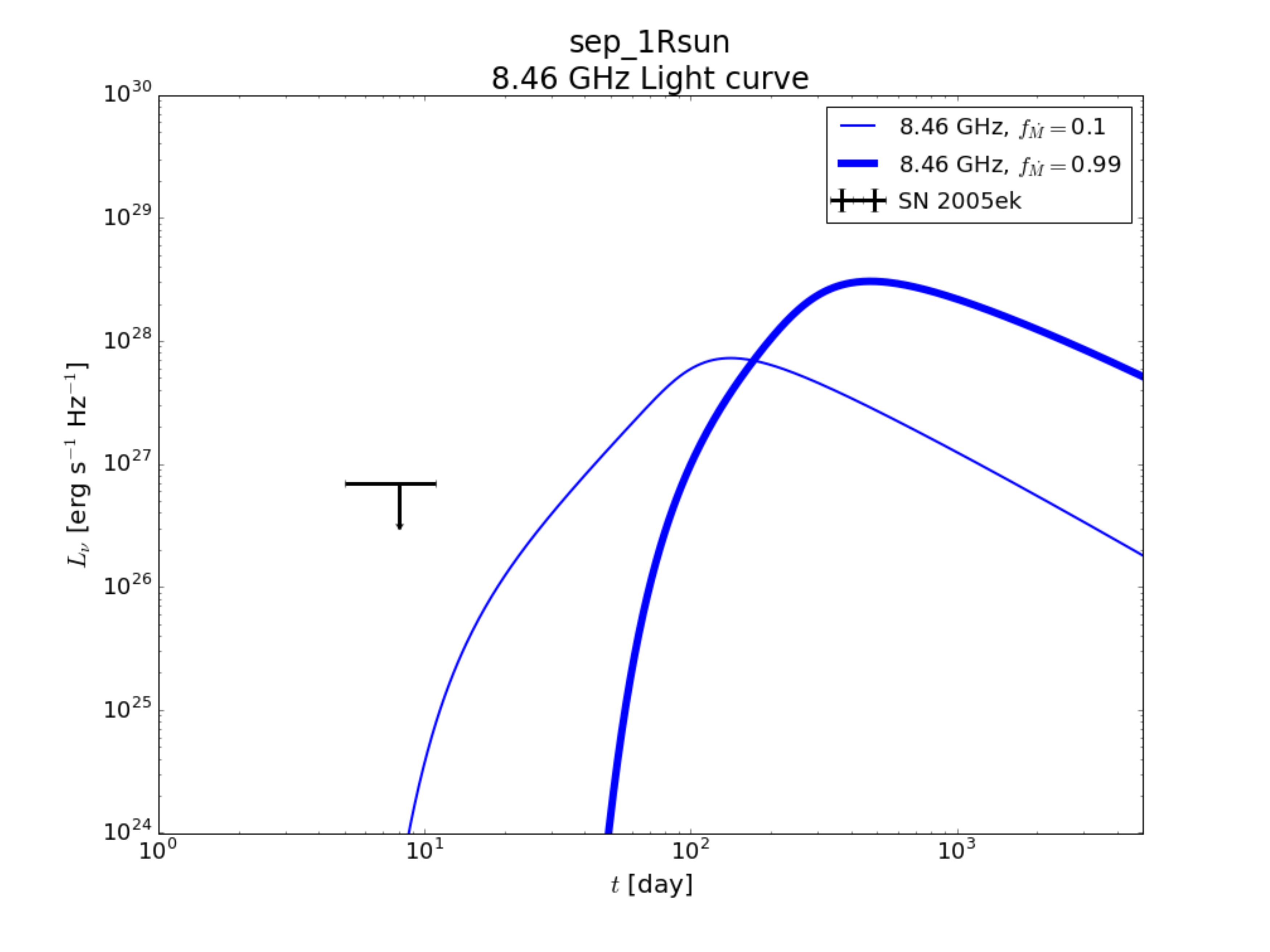}{0.4\textwidth}{(a)}
\fig{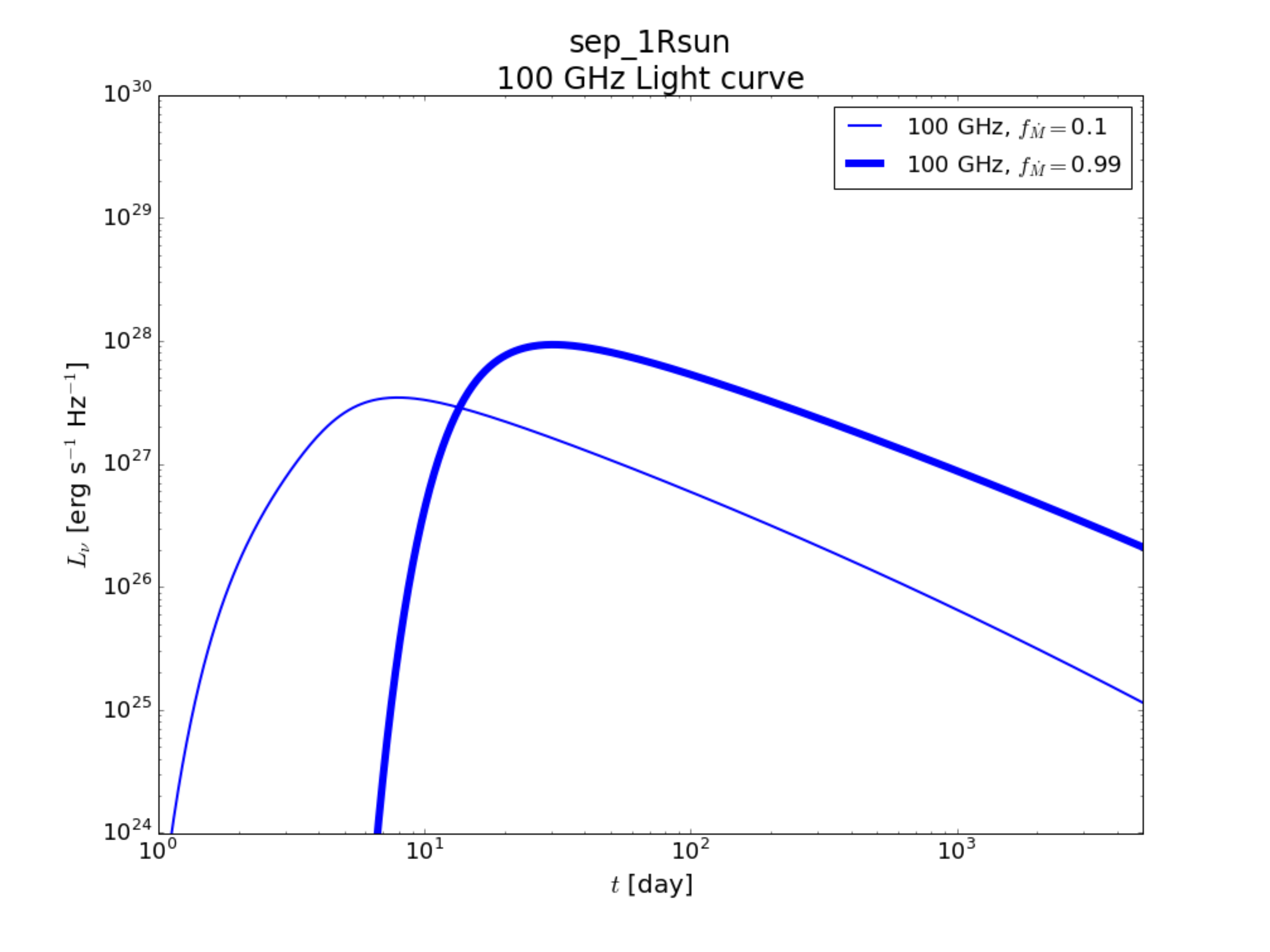}{0.4\textwidth}{(b)}
}
\gridline{
\fig{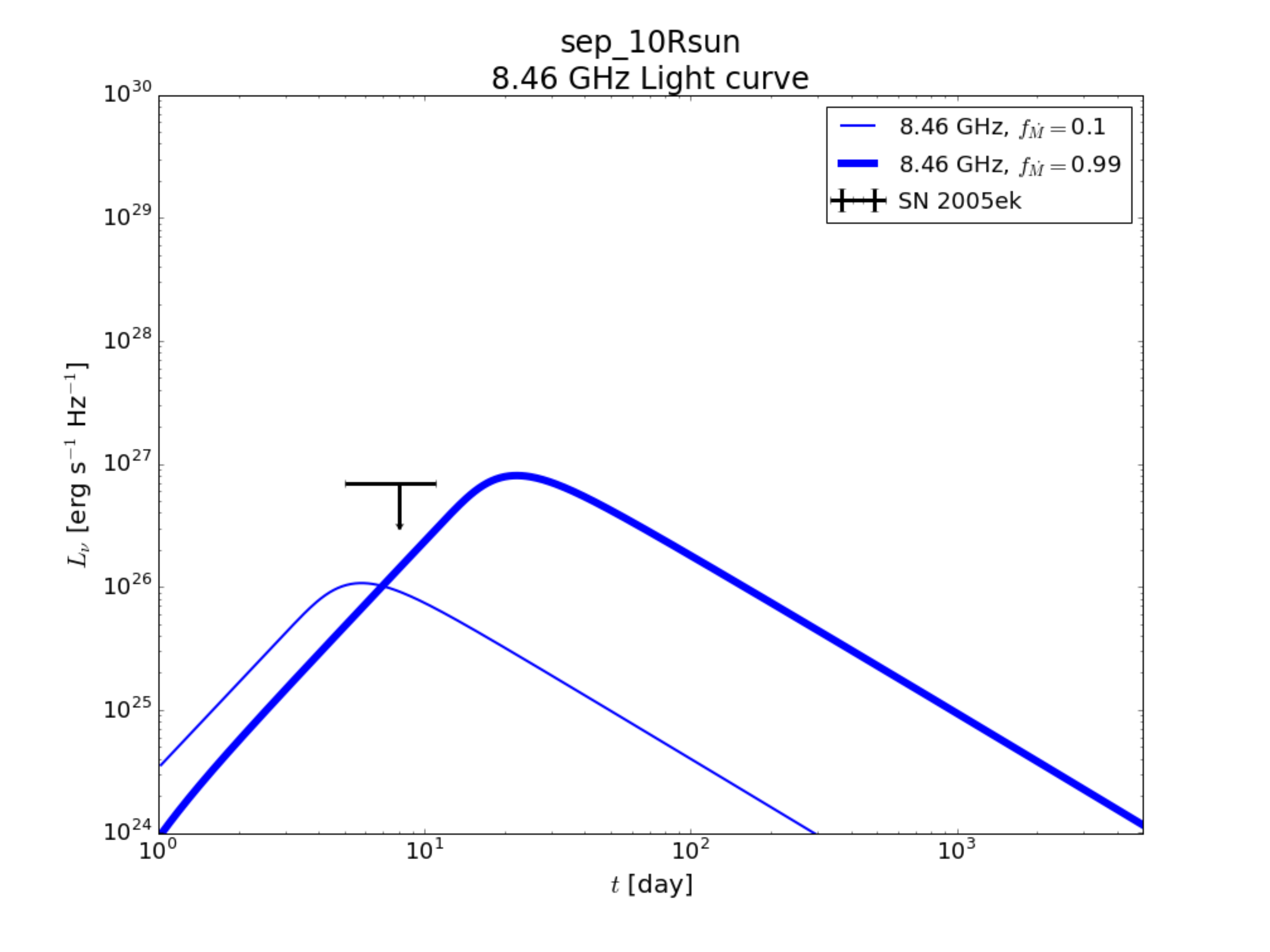}{0.4\textwidth}{(c)}
\fig{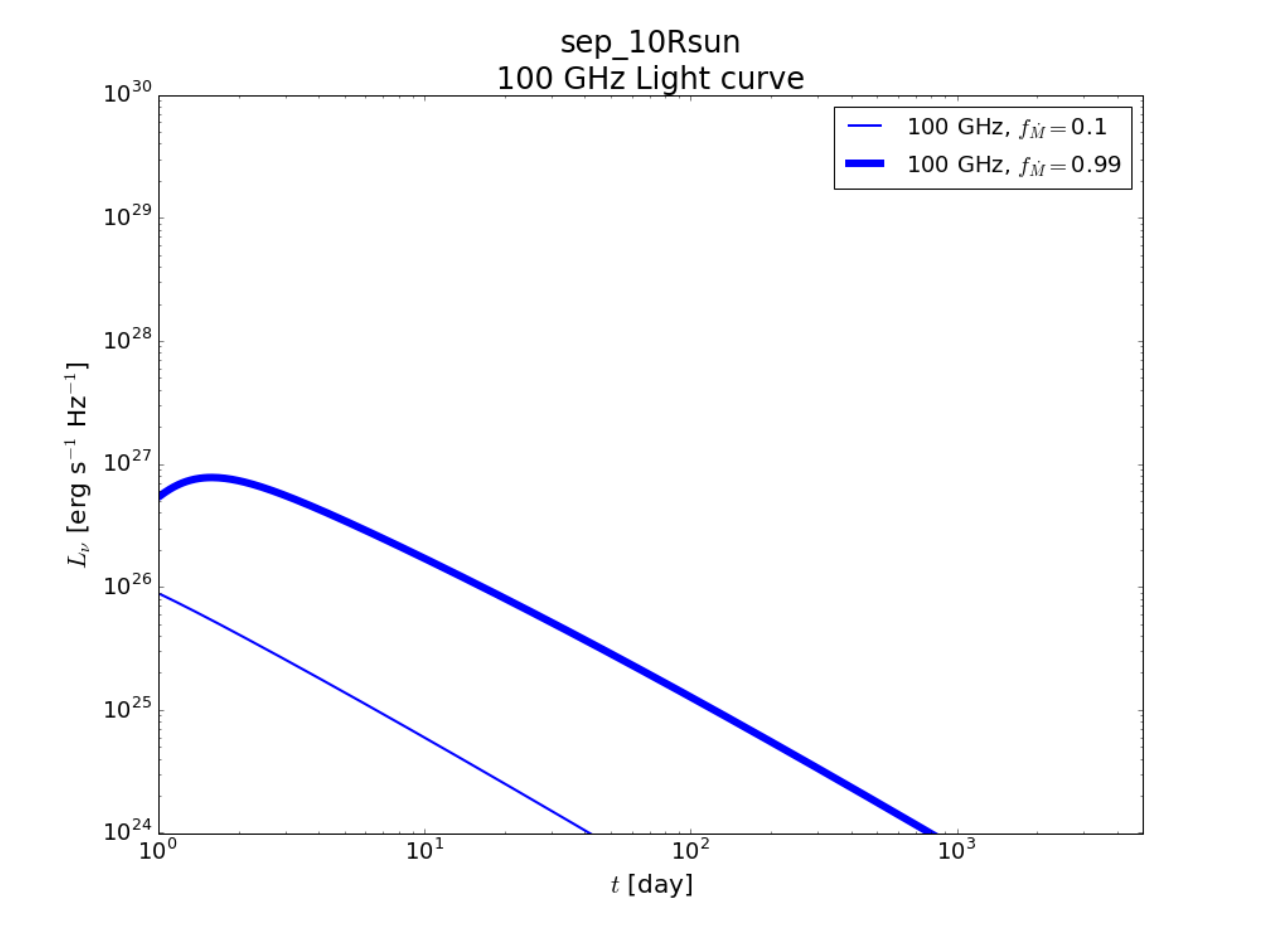}{0.4\textwidth}{(d)}
}
\caption{Examples of the synthesized radio light curves for the models shown in Table 1. The top and bottom panels are for the different models (sep\_1Rsun and sep\_10Rsun). The left and right panels are centimeter (8.46 GHz) and millimeter (100 GHz) ranges, respectively. The thickness of the lines shows the difference in $f_{\dot{M}}$; $f_{\dot{M}}=0.10$ (thin blue) and $f_{\dot{M}}=0.99$ (thick blue). The upper limit of the centimeter emission for SN 2005ek is shown by the {black} arrow.}
\label{fig:LCs}
\end{figure*}

\begin{deluxetable*}{ccccc}[ht!]
\tablecaption{Reference models}
\tablehead{
\colhead{model} &
\colhead{$E_{\rm kin}$ [erg]} &
\colhead{$M_{\rm ej}$ $[M_\odot]$} &
\colhead{$\dot{M}_{\rm RLO}$ [$M_\odot$ yr$^{-1}$]\tablenotemark{a}}&
\colhead{comments}
}
\startdata
sep\_1Rsun & $10^{50}$ & 0.10 & $1.1\times 10^{-2}$ & Model for SN 2005ek \citep{2013ApJ...778L..23T}\\
sep\_10Rsun & $10^{50}$ & {0.06} & ${3.8}\times 10^{-5}$ & - \\ 
\enddata
\end{deluxetable*}
Figure \ref{fig:LCs} shows the light curves at the frequency 8.46 and 100 GHz with $f_{\dot{M}} = 0.10$ and $0.99$. The models shown here are described in Table 1. {These models are selected as representative cases having $M_{\rm ej} \sim 0.1M_\odot$ to be consistent with the ejecta mass estimated for the ultra-stripped SN candidates SN 2005ek and iPTF 14gqr, through their rapid evolutions in the optical light curves (Sections \ref{sec:progenitors} and \ref{sec:models}).
The explosion energy $E_{\rm kin} = 10^{50}$ erg is predicted by the theoretical simulation \citep{2015MNRAS.454.3073S}, which is also consistent with those estimated for SN 2005ek and iPTF 14gqr \citep{2013ApJ...778L..23T, 2018Sci...362..201D}.} 

Qualitatively, the radio emission from SNe at higher frequency becomes transparent to the CSM in the earlier epoch. This trend can be seen in all of the models. For example, in the model sep\_1Rsun, where the dense CSM is distributed ($\dot{M}_{\rm CSM} \gtrsim 10^{-3} \ M_\odot$yr$^{-1}$), the synchrotron emission at 8.46 GHz is damped by the strong SSA and FFA in the first 10 days, while in the late epochs (100 - 1000 days), it shows a high luminosity. This behavior is commonly seen in the observed radio emission from Type IIn SNe \citep[e.g.,][]{1998ApJ...499..810C}. 
On the other hand, the emission at 100 GHz is peaked within 1 month, followed by the optically thin, decaying emission in 100 - 1000 days. 

In the model sep\_10Rsun, the peak date of the radio luminosity at 8.46 GHz is at 10 - 100 days, which is {similar to} the observed radio {peak dates for} typical Type Ib/Ic SNe \citep[see e.g.,][]{2014ApJ...797..107M, 2019ApJ...883..147T}. However, the maximum luminosity is smaller than {those} of the typical Type Ib/Ic SNe, because of the low explosion energy {of the ultra-stripped SNe (see also Section \ref{sec:confinedCSM})}. The 100 GHz emission is {similarly weak} ($L_\nu \sim 10^{26}$ erg s$^{-1}$ Hz$^{-1}$) {which is peaked} at $t \lesssim 10$ days. 

\subsection{Maximum Luminosities}\label{sec:Lmax}
\begin{figure*}[ht!]
\gridline{
\fig{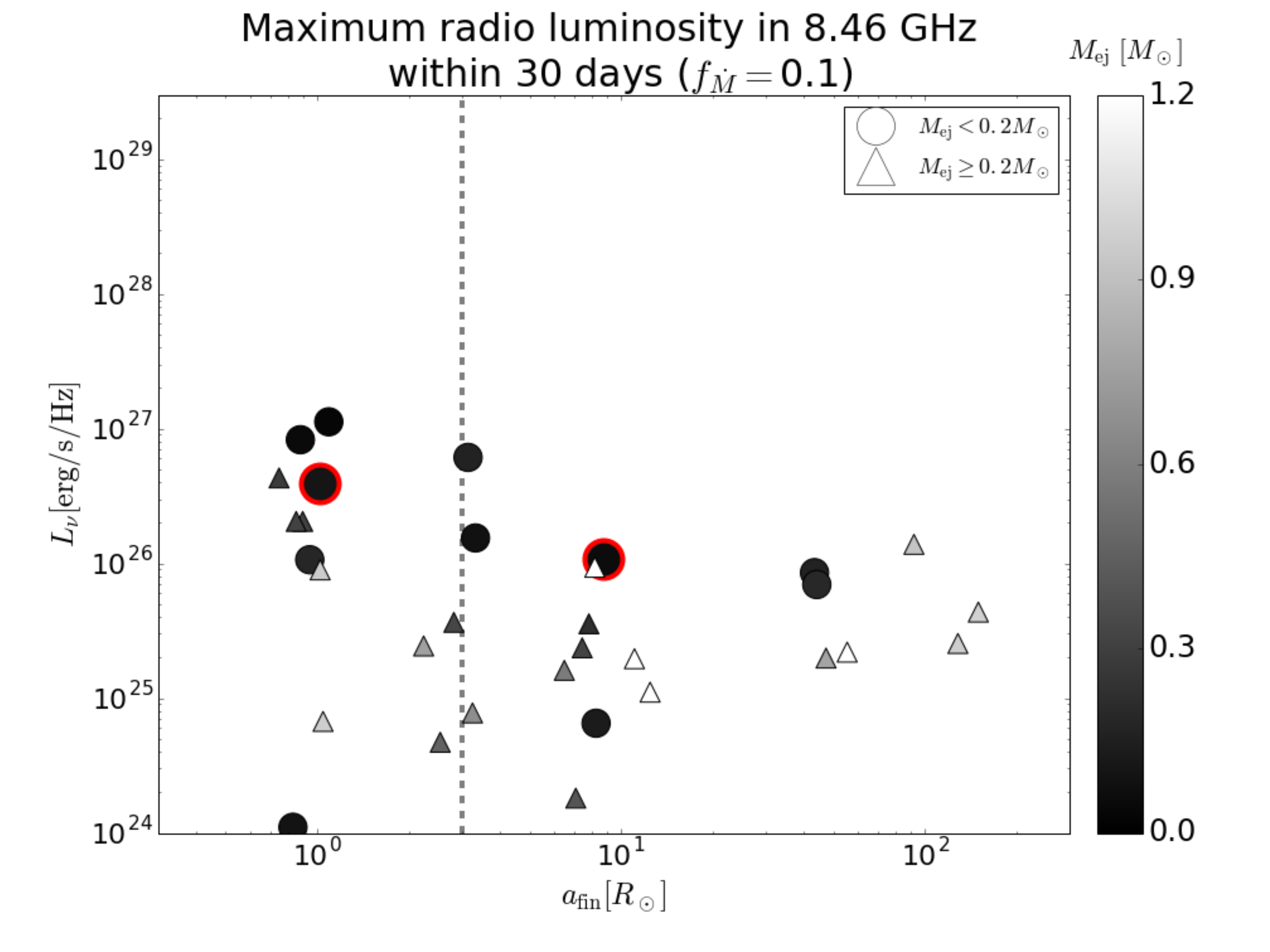}{0.4\textwidth}{(a)}
\fig{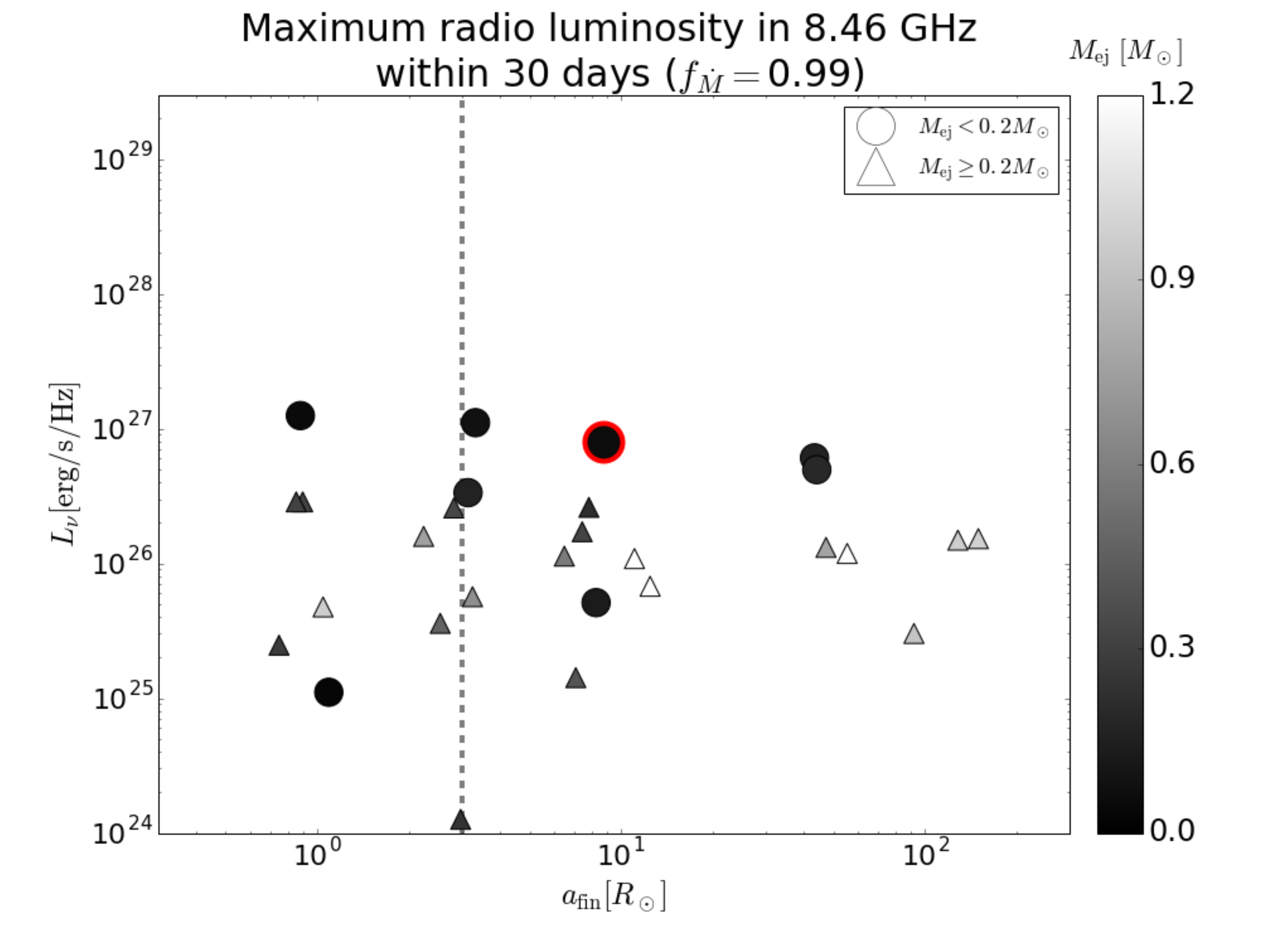}{0.4\textwidth}{(b)}
}
\gridline{
\fig{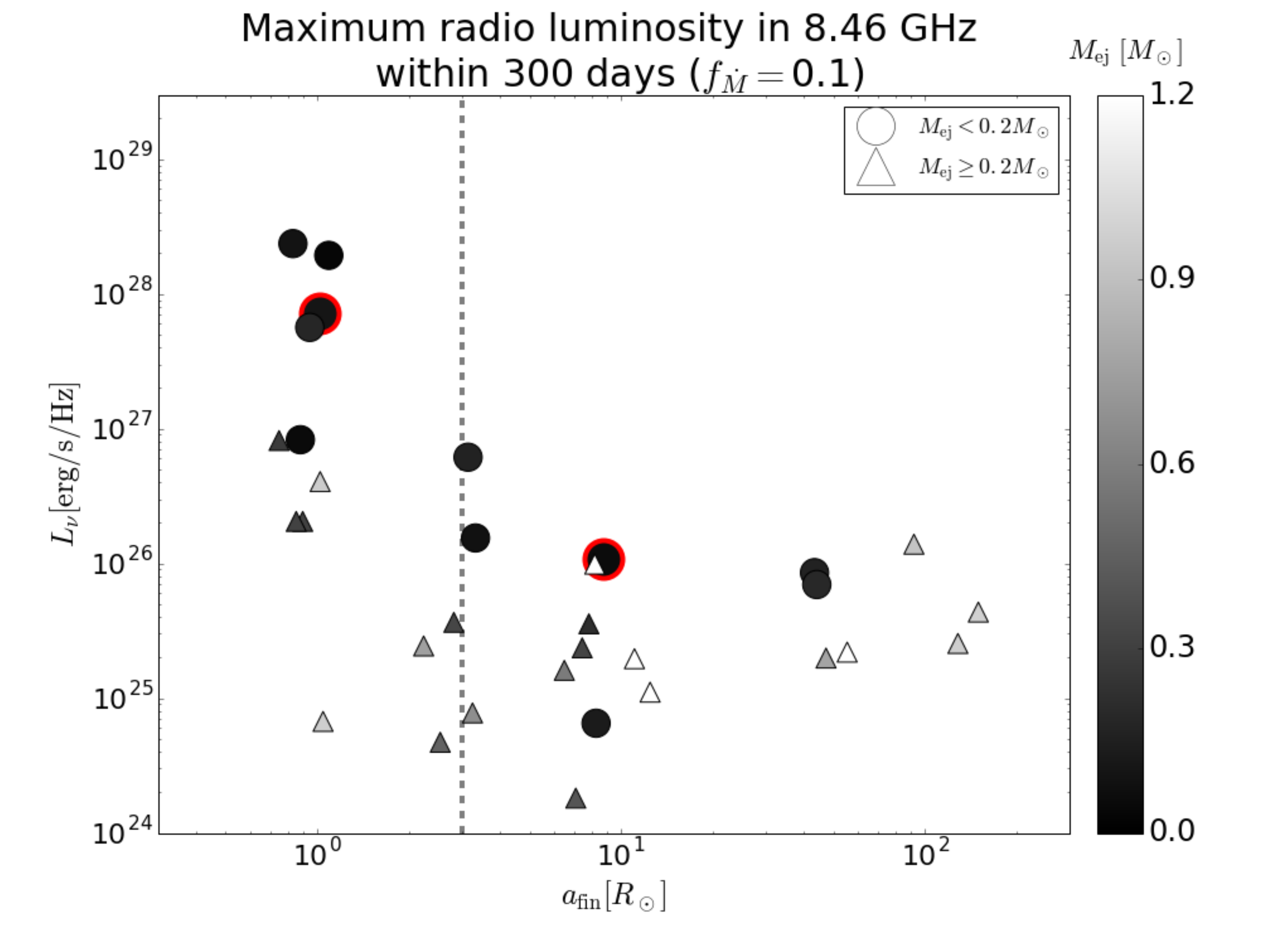}{0.4\textwidth}{(c)}
\fig{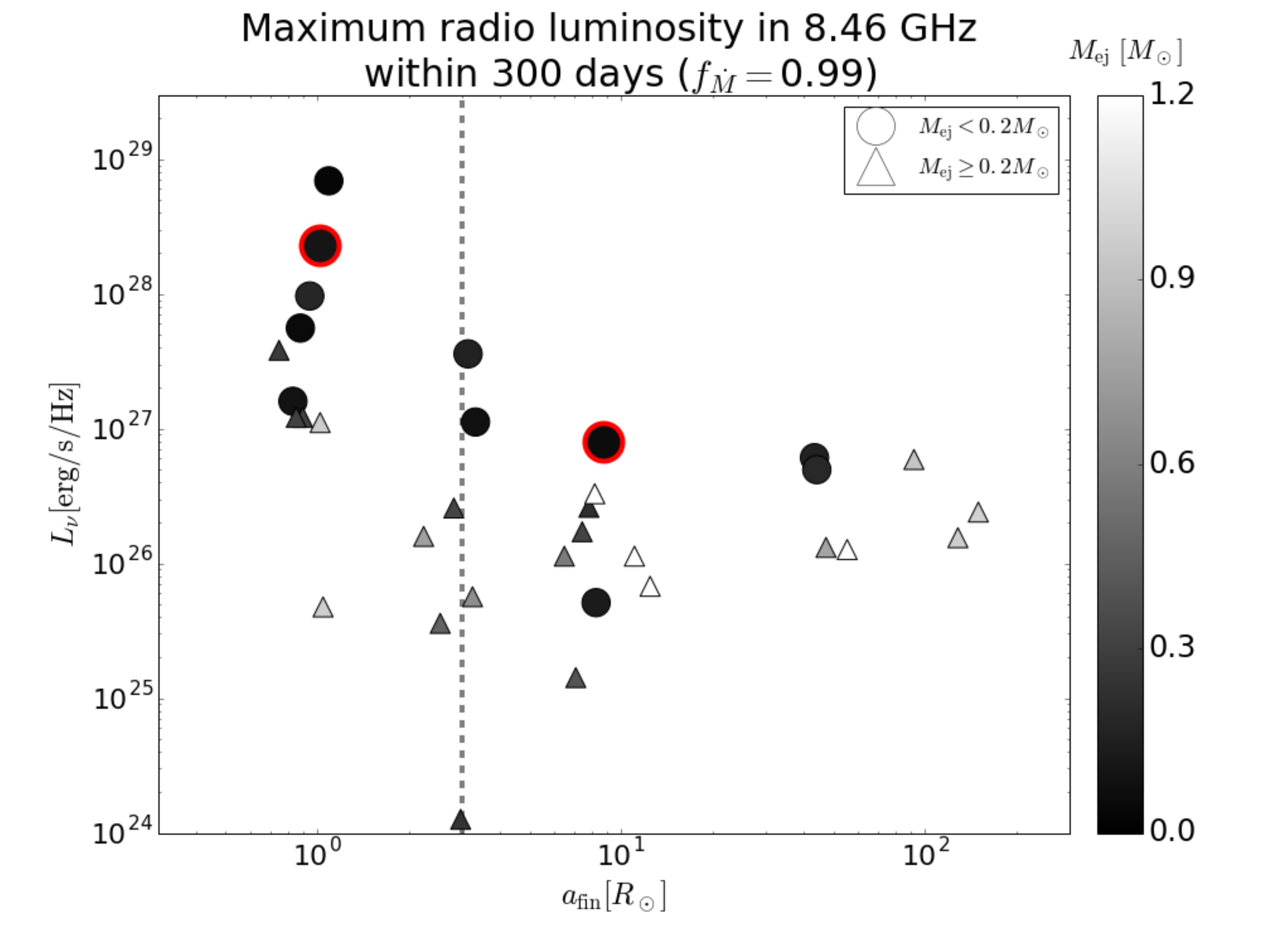}{0.4\textwidth}{(d)}
}
\gridline{
\fig{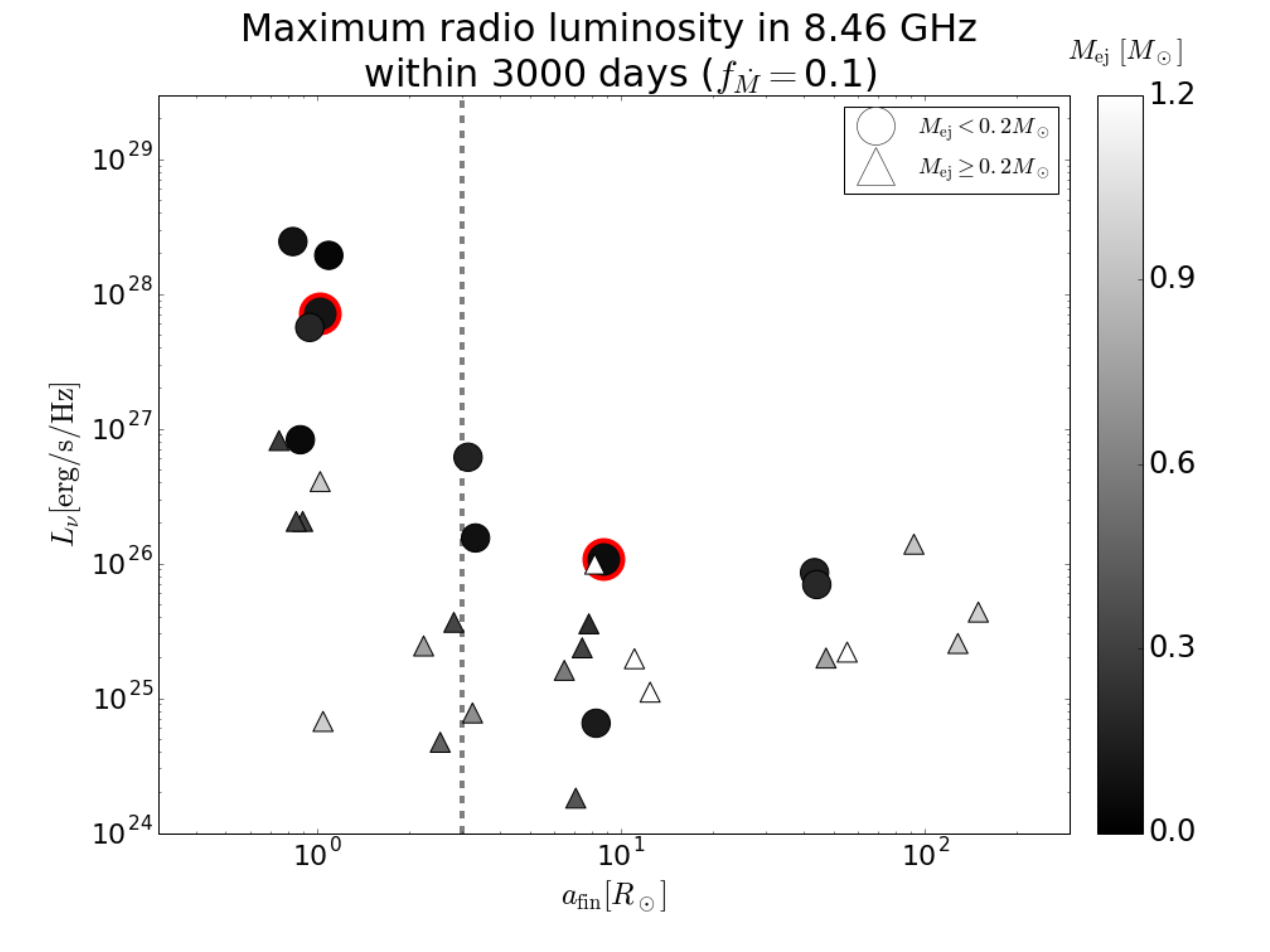}{0.4\textwidth}{(e)}
\fig{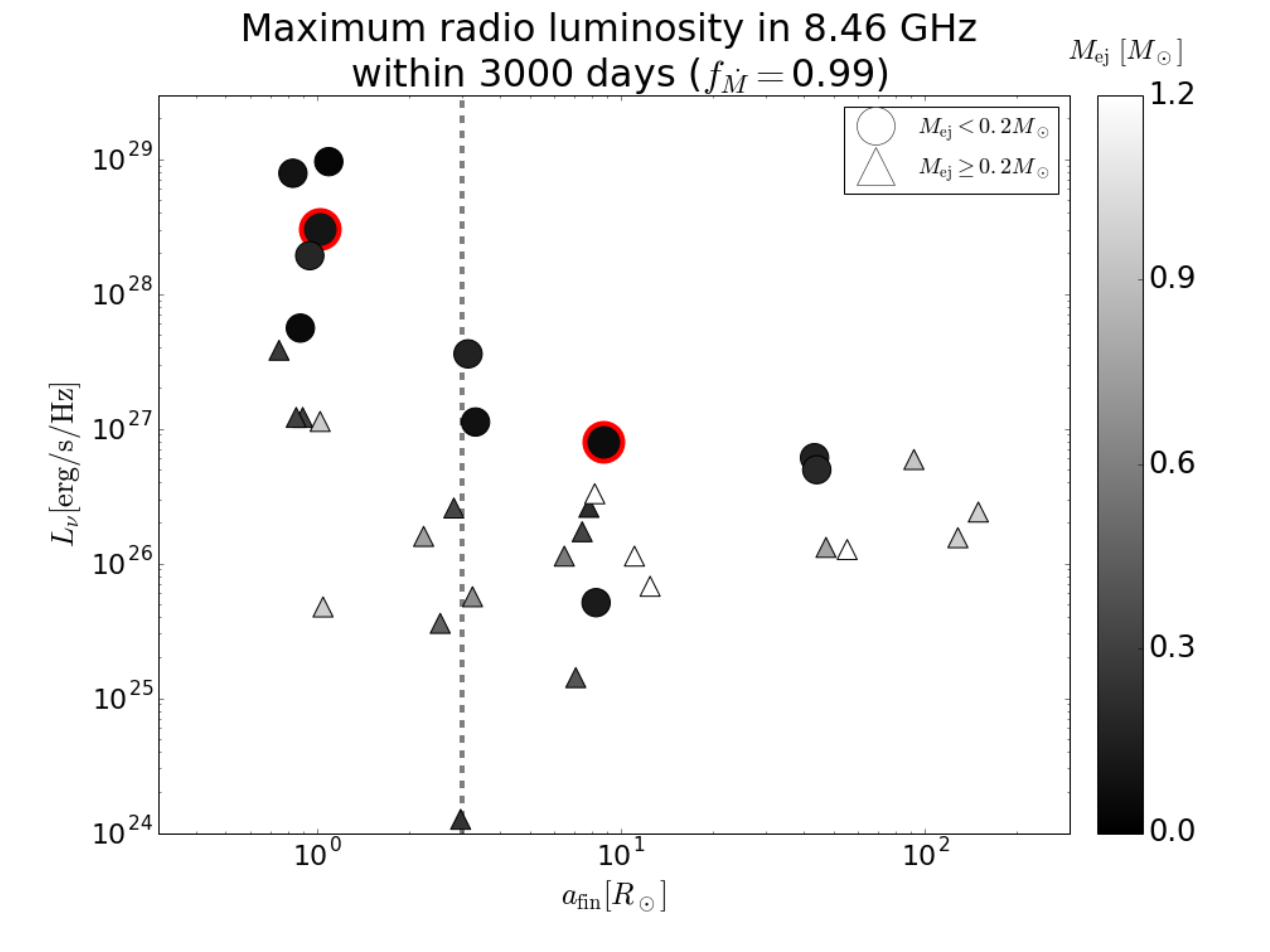}{0.4\textwidth}{(f)}
}
\caption{{Dependences of the radio maximum luminosity on} the final separation, for different time windows since the explosion (within the first 30 days, 300 days, and 3000 days from the top to bottom). The frequency is set at 8.46 GHz in these figures. The left panels are for $f_{\dot{M}} = 0.10$, while the right ones are for $f_{\dot{M}} = 0.99$.}
\label{fig:cm}
\end{figure*}

Figure \ref{fig:cm} shows the distribution of the maximum luminosities of the centimeter emission (8.46 GHz) for various models, measured within fixed time interval of the first 30, 300, and 3000 days, as a function of the final binary separation. The maximum luminosities in the centimeter range within the first 30 days do not show a characteristic difference among all {of} the binary evolution models. However, if we extend the time-window to 300 or 3000 days, we can observe strong centimeter emissions ($ \gtrsim 10^{28}$ erg s$^{-1}$ Hz$^{-1}$) from {some models with $a_{\rm fin} \sim 1R_\odot$}. This strong radio signal can be robust {diagnostics} {for} the dense CSM around the progenitor; a large amount of the helium layer of the progenitor has been stripped {away, and the small binary separation is responsible for this strong envelope stripping.} 

\begin{figure*}[ht!]
\gridline{
\fig{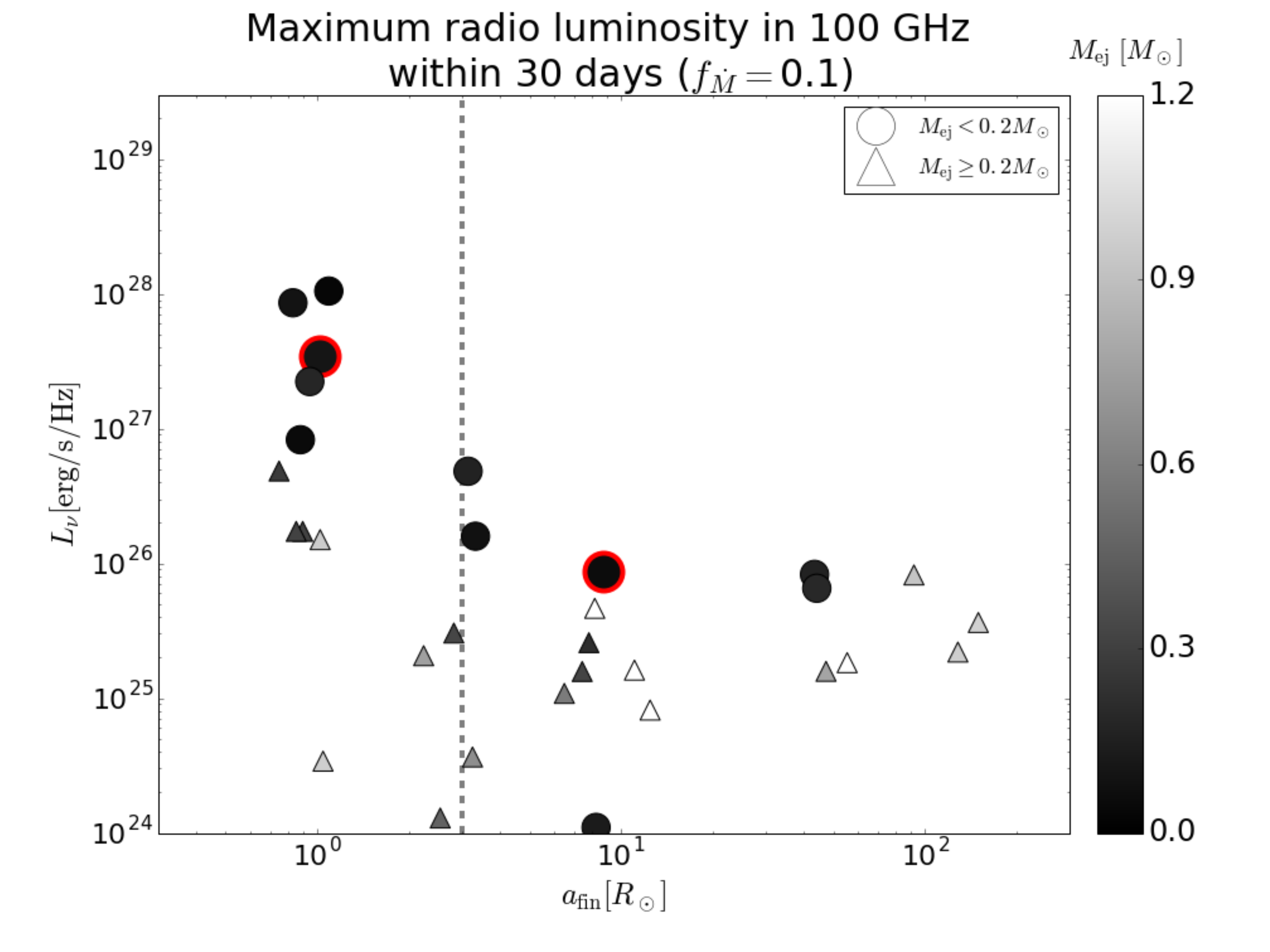}{0.4\textwidth}{(a)}
\fig{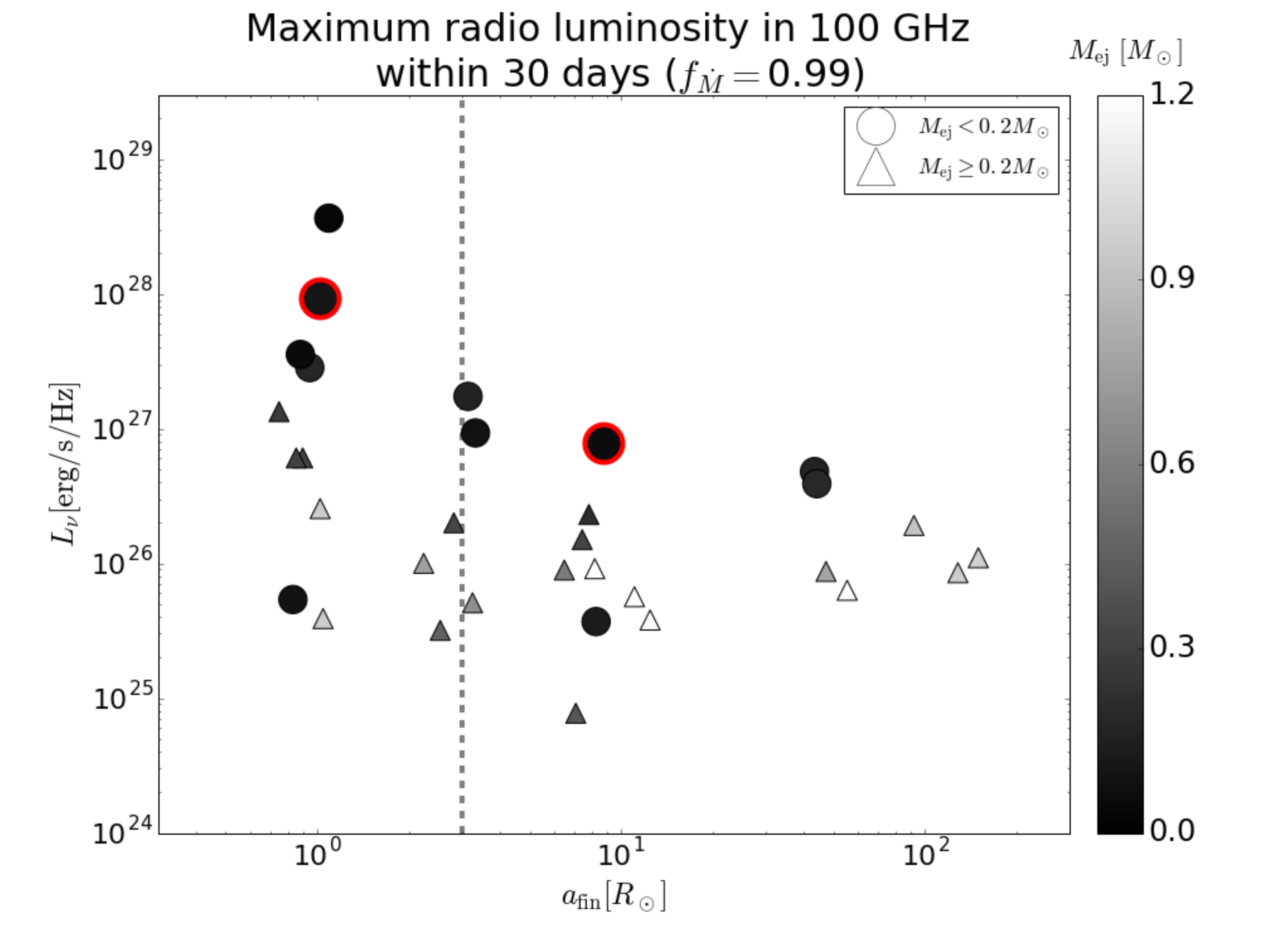}{0.4\textwidth}{(b)}
}
\gridline{
\fig{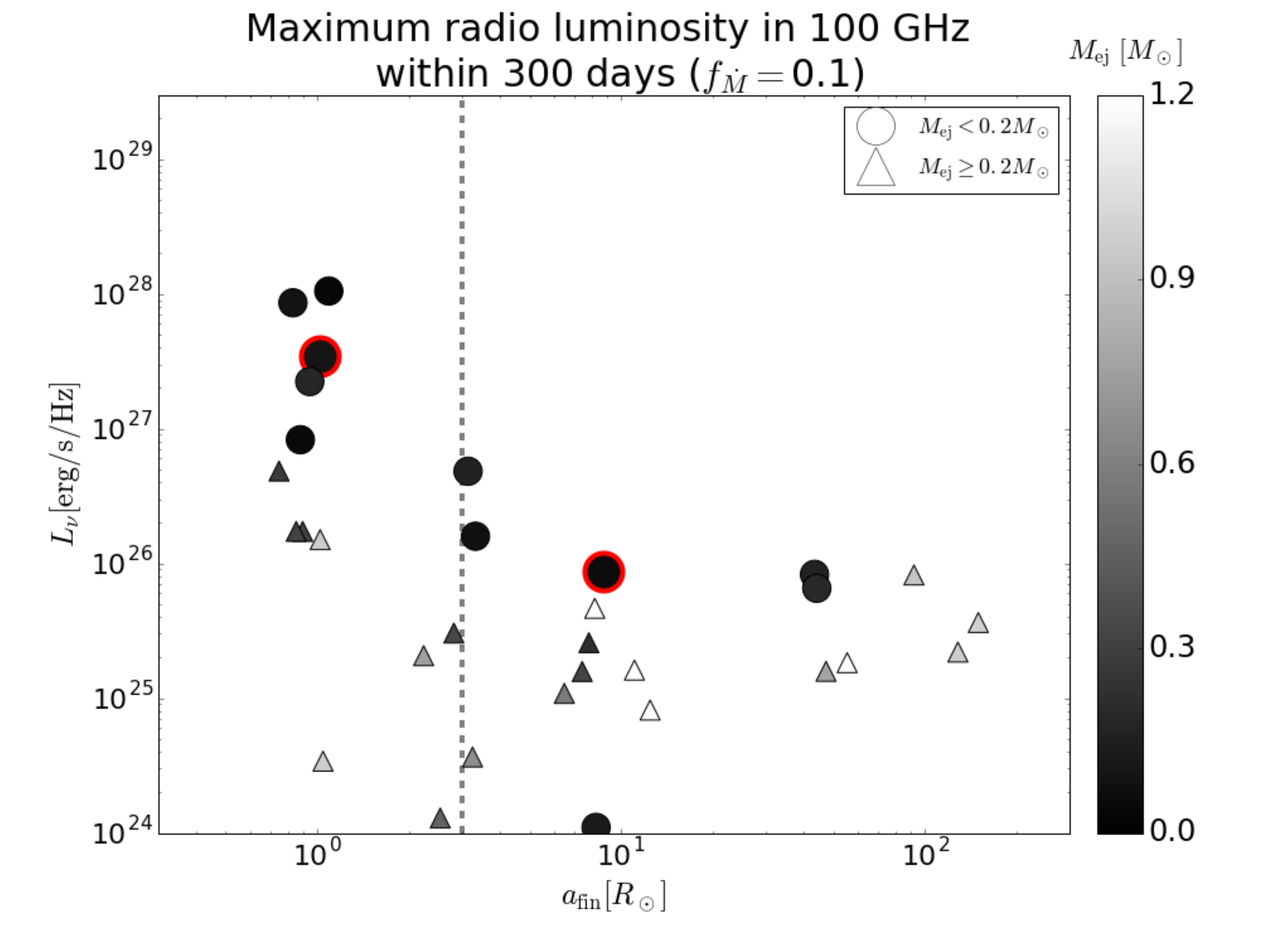}{0.4\textwidth}{(c)}
\fig{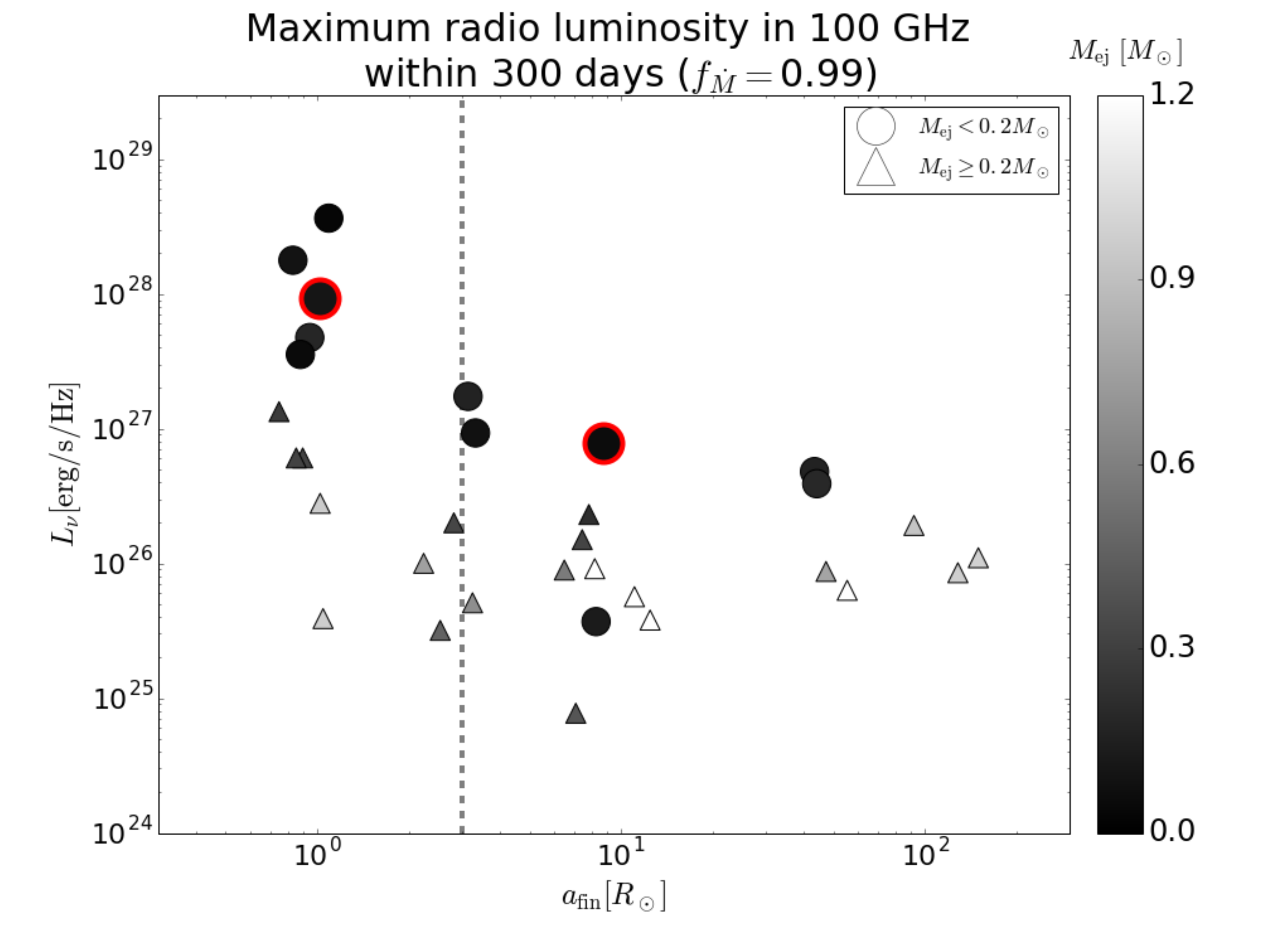}{0.4\textwidth}{(d)}
}
\gridline{
\fig{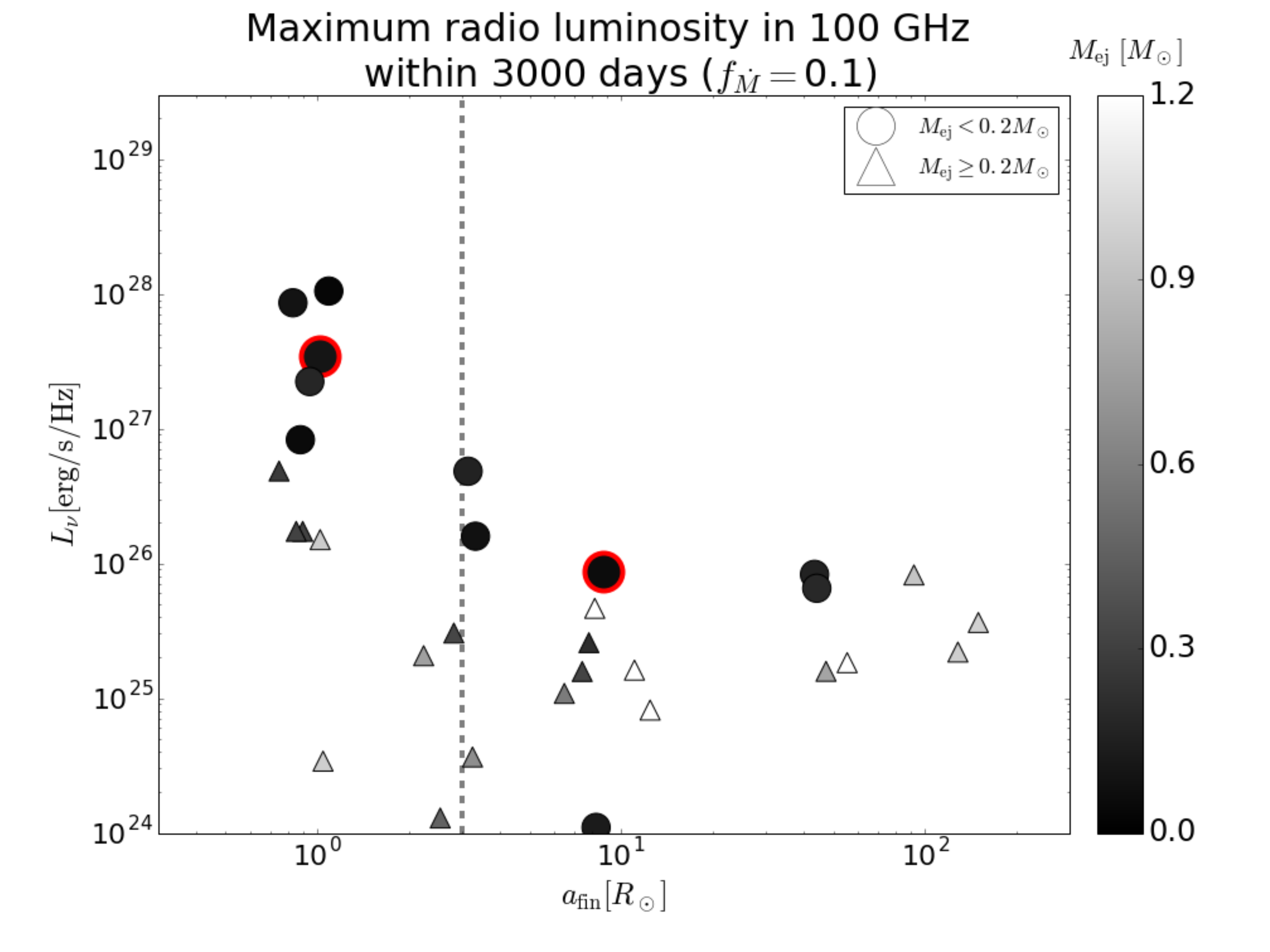}{0.4\textwidth}{(e)}
\fig{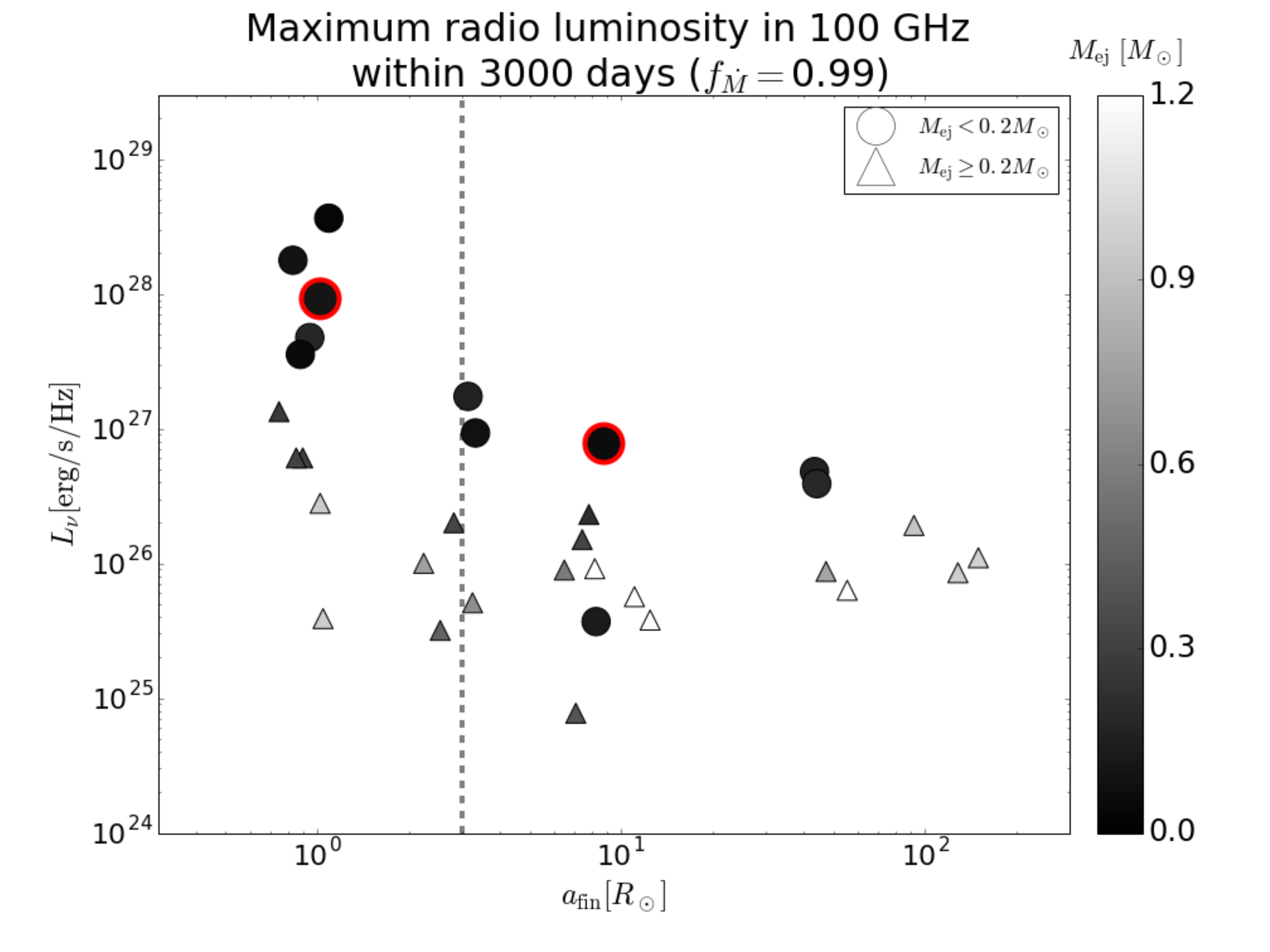}{0.4\textwidth}{(f)}
}
\caption{Same as Figure 4, but for 100 GHz.}
\label{fig:mm}
\end{figure*}

Figure \ref{fig:mm} shows the same result as Figure \ref{fig:cm}, but for the millimeter emission (100 GHz). The millimeter emissions are enhanced in some models with $a_{\rm fin} \sim 1R_\odot$ already within 30 days. {The bright early-phase millimeter signals, {as well as the bright late-phase centimeter signals}, are useful as indicators of the small binary separation}. We note that the maximum luminosity in the millimeter range is reached within 30 days for all of the models, followed by an optically thin, decaying emission in the late epoch (see Figure \ref{fig:LCs}). 
These features lead to the optimized strategy to detect the radio signals from ultra-stripped SNe; either a long-term monitoring in the {centimeter} range or a quick follow-up in the millimeter {range}, or a combination of both.

{If the observed maximum luminosity (or upper limit) {is} $L_\nu \lesssim 10^{26}$ erg s$^{-1}$ Hz$^{-1}$ (both in centimeter range and in millimeter range), the interpretation will not be straightforward. Many models with a range of the final binary separation can lead to $L_\nu \sim 10^{26}$ erg s$^{-1}$ Hz$^{-1}$. However, this {behavior} is largely driven by the models with the large ejecta mass; if we focus on the models with the small ejecta mass ($M_{\rm ej} < 0.2M_\odot$), there is a tendency for $L_\nu$ to decrease as $a_{\rm fin}$ increases (see e.g., (f) in Figure \ref{fig:cm} or (b) in Figure \ref{fig:mm}). This trend results from the correlation between $\dot{M}_{\rm RLO}$ and $a_{\rm fin}$ discussed in Section \ref{sec:progenitors}. The ejecta mass can be estimated by the optical data. Therefore, the combination of the observational data in the radio and optical ranges might be a useful indicator of an ultra-stripped SN forming a DNS binary which {will} not merge within the cosmic time, in case the radio signal is weak.}

{Our findings on the relation between the observed radio luminosity and the binary evolution scenario are summarized as follows.} Strong signals around 300 - 3000 days {in the centimeter range or within 300 days in the millimeter range indicate} large $\dot{M}_{\rm RLO}$ due to small separation. {For such a system,} the DNS binary left after the ultra-stripped SN event will make a coalescence within the cosmic age. On the other hand, a low radio luminosity implies small $\dot{M}_{\rm RLO}$, {which can be realized for a wide range of the binary separation. However, once small ejecta mass is inferred by the optical data, it {is} suggested that the binary system has large separation {and} the remnant DNS binary {will not} merge within the cosmic age.} 

\section{Discussions}\label{sec:discussions}
\subsection{Candidates for the ultra-stripped SNe and their radio observations}\label{sec:obs_table}
\begin{deluxetable*}{c|cccc}[ht!]
\tablecaption{Radio observations of the candidates for ultra-stripped SNe}
\tablehead{
\colhead{Object} &
\colhead{Distance [Mpc]} &
\colhead{$t_{\rm obs}$ [day]}\tablenotemark{a} &
\colhead{$\nu_{\rm obs}$ [GHz]}\tablenotemark{b}&
\colhead{upper limit of $L_\nu$ [erg s$^{-1}$ Hz$^{-1}$]}
}
\startdata
SN 2005ek & 67 & $5 - 11$ & 8.46 & $7\times10^{26}$ \\
iPTF 14gqr & 284.5& $\sim 1$ & 15 & $5.6\times10^{27}$ \\
iPTF 14gqr & - & $\sim 1.7$ & 6.1& $1.12\times10^{27}$ \\
iPTF 14gqr & - & $\sim 1.7$ & 22 & $1.13\times10^{27}$ \\
iPTF 14gqr & - & $\sim 11$ & 6.1 & $1.26\times10^{27}$ \\
iPTF 14gqr & - & $\sim 11$ & 22 & $1.50\times10^{27}$ \\
\enddata
References: 
SN 2005ek : \cite{2013ApJ...774...58D}, 
iPTF 14gqr : \cite{2018Sci...362..201D}
\tablenotetext{a}{Observed epoch since the explosion.}
\tablenotetext{b}{Observational frequency.}
\label{table:observations}
\end{deluxetable*}

Table \ref{table:observations} summarizes the constraints {on radio luminosities of} the ultra-stripped SN candidates (SN 2005ek and iPTF 2014gqr). All of the observations were conducted in the centimeter range within 10 days since the explosion, although the observed epoch ($t_{\rm obs}$) of SN 2005ek has an uncertainty due to the unknown explosion date \citep[see][]{2013ApJ...774...58D}. Non-detections are reported {in} all {of the} cases. The upper limits for the radio luminosity per unit frequency are {given as} $L_\nu \lesssim 10^{27}$ erg s$^{-1}$ Hz$^{-1}$. While these observations in principle allow us to investigate the nature of the CSM, these available upper limits are not {deep} enough to be a strong constraint (see Figure \ref{fig:LCs} for SN 2005ek).

\subsection{Strategy in radio follow-up observations}\label{sec:obs_strategy}
\begin{figure*}[ht!]
\gridline{
\fig{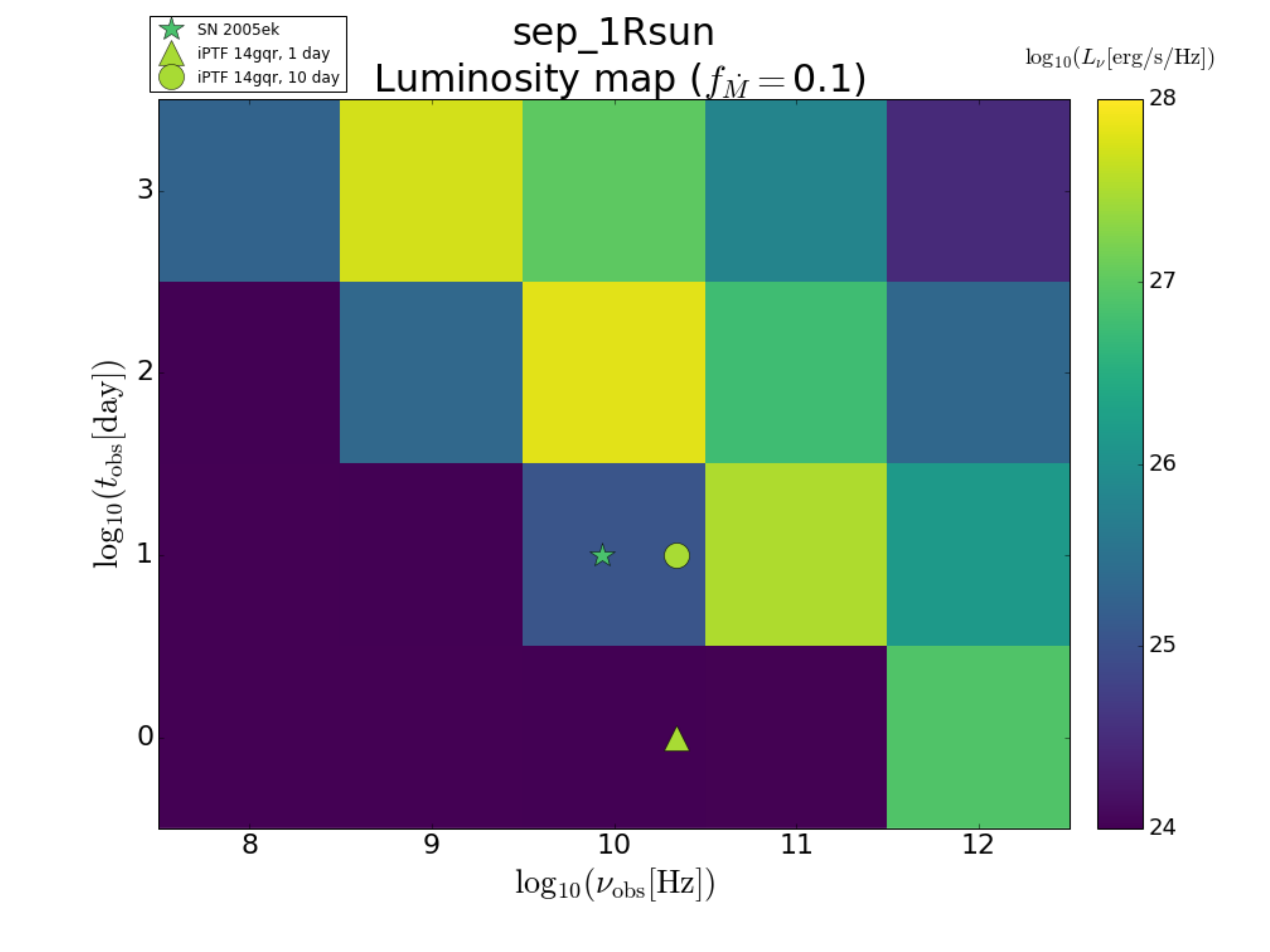}{0.4\textwidth}{(a)}
\fig{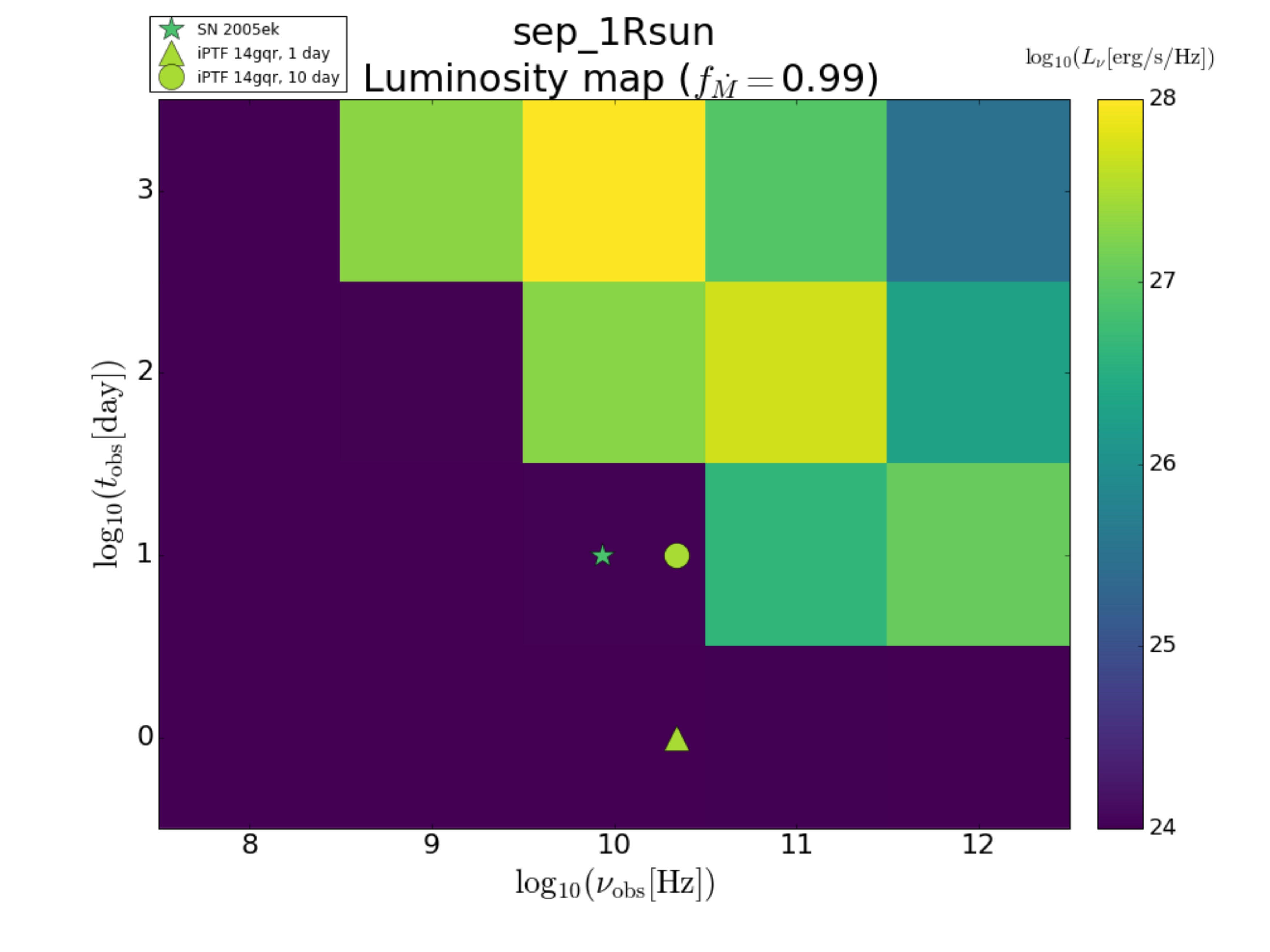}{0.4\textwidth}{(b)}
}
\gridline{
\fig{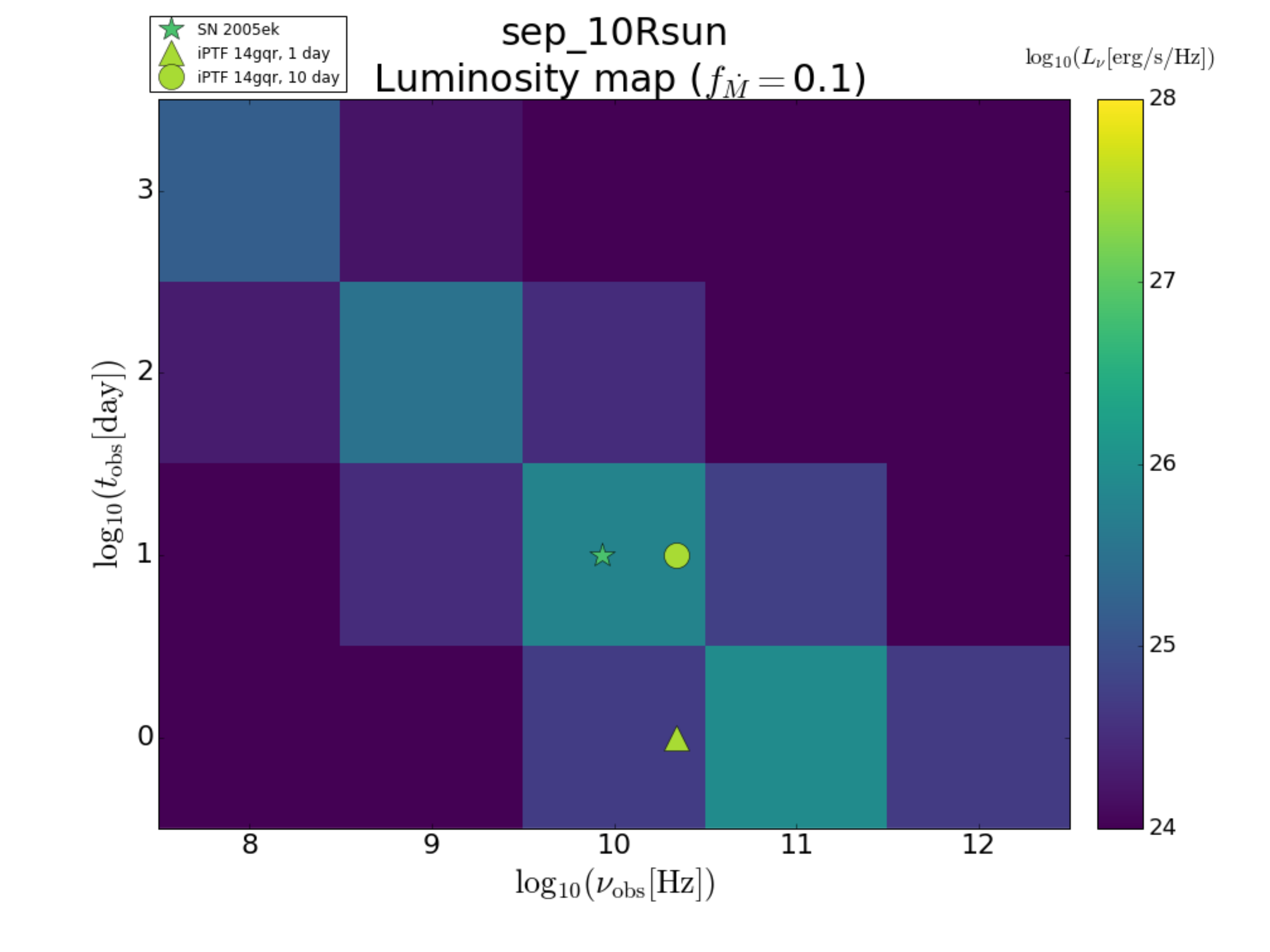}{0.4\textwidth}{(c)}
\fig{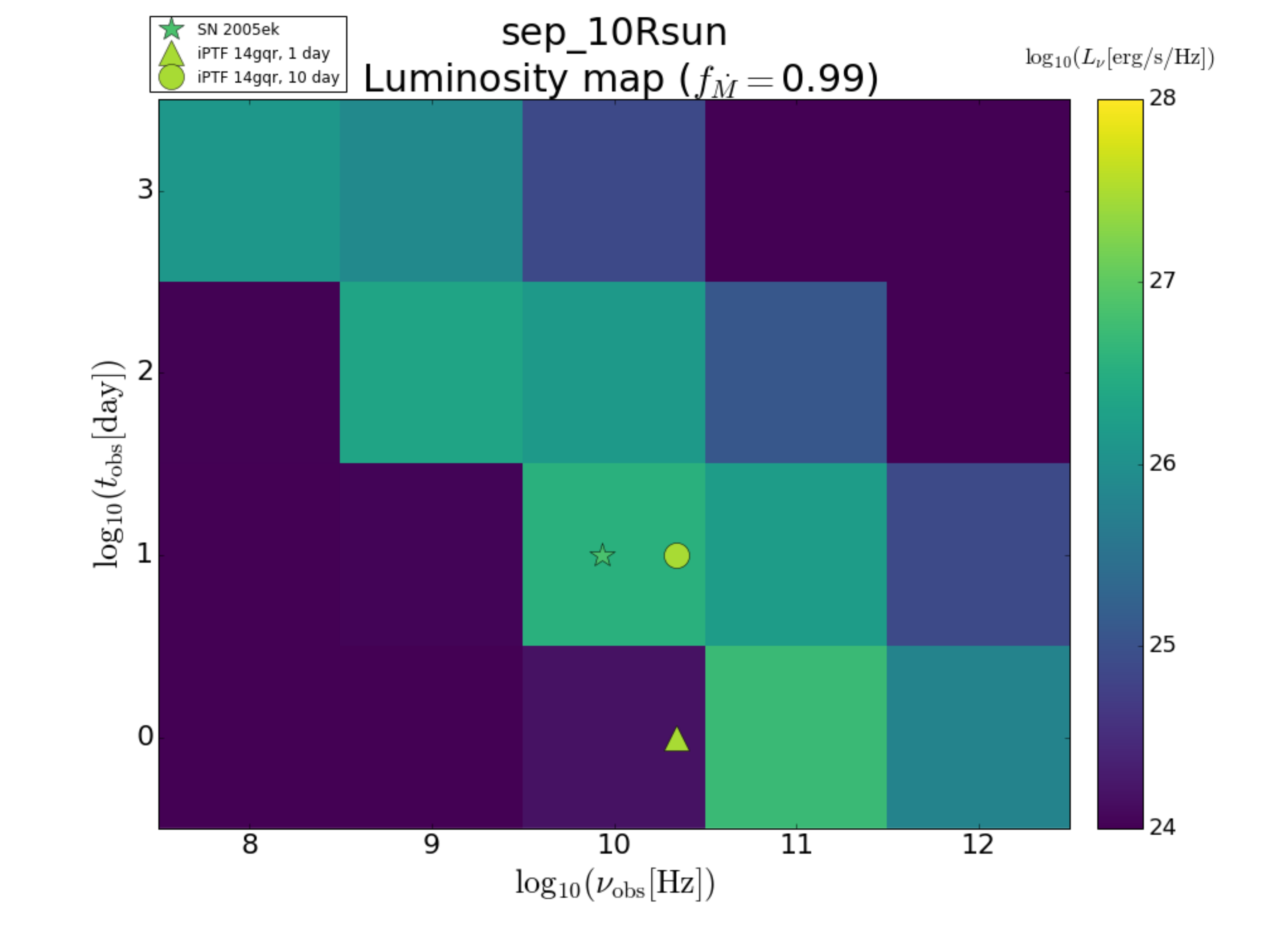}{0.4\textwidth}{(d)}
}
\caption{The radio luminosity for the reference models {shown by the different colors}, as functions of the epoch and frequency. The x-axis is the frequency in a {logarithmic scale}, while the y-axis is the epoch in a {logarithmic scale}. {Shown} here are the {models} sep\_1Rsun (Top) and sep\_10Rsun (Bottom). In the left panels $f_{\dot{M}}=0.1$ is used, while in the right panels $f_{\dot{M}}=0.99$. The points show a combination of the epoch and frequency in the past observations (Table \ref{table:observations}).}
\label{fig:colors}
\end{figure*}

Figure \ref{fig:colors} summarizes the radio luminosity as functions of the observational epoch and frequency for the {ultra-stripped SN} models shown in Table 1. For the model sep\_1Rsun, {the} centimeter emission after 100 days {or} the millimeter emission around 10-100 days {provides} the optimized windows. For the model sep\_10Rsun, {the centimeter emission at 10-100 days or millimeter emission within 10 days} would become the best tracer of the CSM, although {even for these windows} the peak luminosity is less than $10^{27}$ erg s$^{-1}$ Hz$^{-1}$. 

The previous radio observations for the ultra-stripped SN {candidates were} conducted {in the centimeter range within 10 days, during which the absorption effect is still strong (see Table \ref{table:observations})}. In such an early phase, the signal will be damped by the SSA and the FFA. The rapid observation in the centimeter range is not suitable as the {diagnostics} of the binary separation. Hence, it is necessary to continue the centimeter observation {until} $t\sim 100-1000$ days. On the other hand, the millimeter emission is enhanced around 10 days. We suggest that a rapid millimeter follow-up observation can be a potential tracer of the nature of the progenitor binary. 

{
\subsection{Contribution from an additional `confined' CSM}\label{sec:confinedCSM}
While our model is constructed based on the RLO mass-transfer history in the ultra-stripped SN models, massive stars may experience an additional mass-loss process for which the origin has not {yet} been clarified. Recently, the possibility of a pre-SN enhanced mass loss, especially for Type II {SN progenitors}, has been proposed \citep[e.g.,][]{2017NatPh..13..510Y, 2018NatAs...2..808F}.  It has {also} been reported that the progenitor of iPTF 14gqr {was} surrounded by the dense CSM existing up to $\sim 10^{15}$ cm \citep{2018Sci...362..201D}. {In this section, we show that} it is possible to distinguish the {potential} radio signal contributed by the confined CSM from that created by the RLO mass transfer associated with the {evolution of the ultra-stripped SN progenitor binary.}

\begin{figure*}[ht!]
\gridline{
\fig{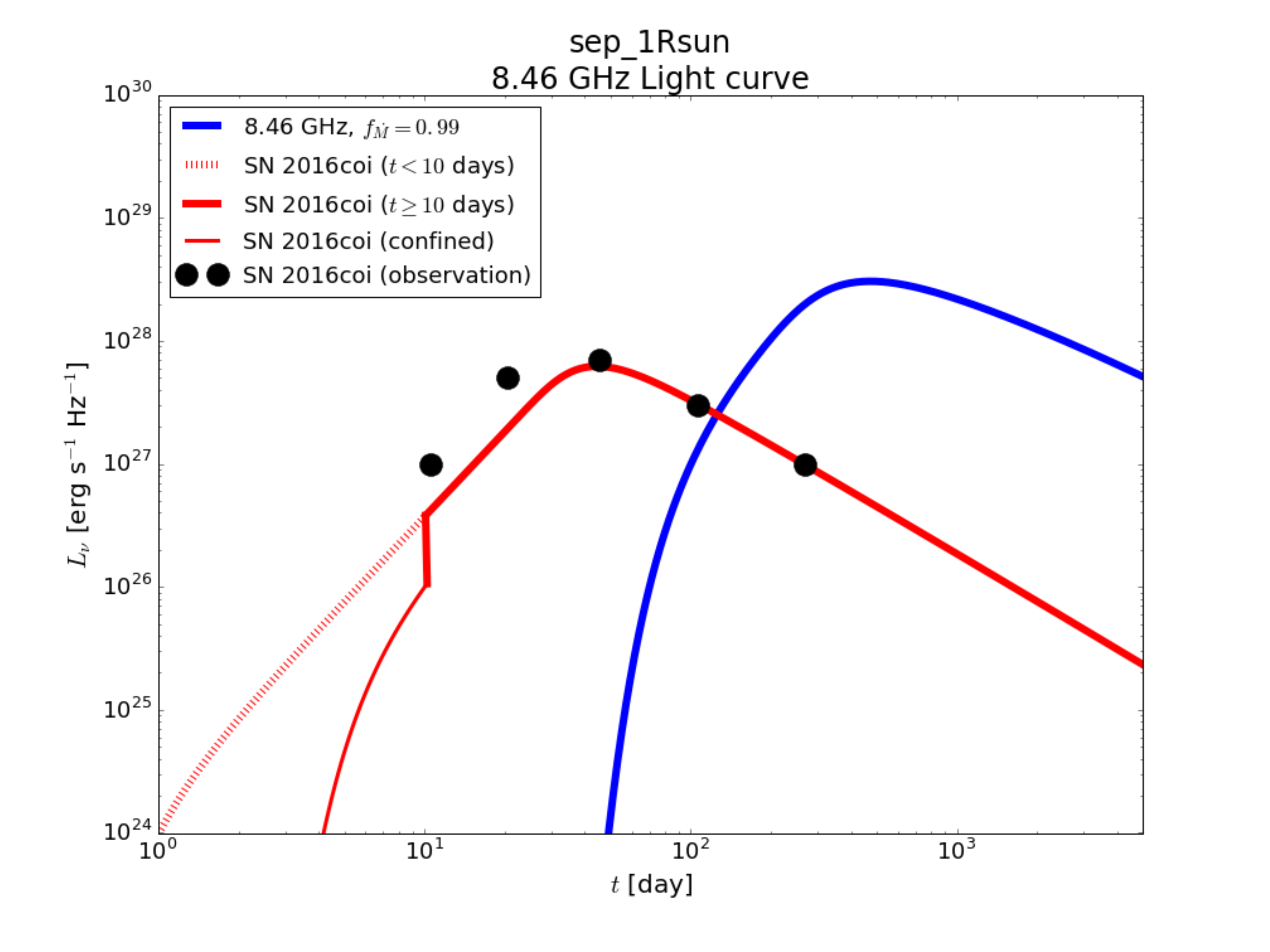}{0.4\textwidth}{(a)}
\fig{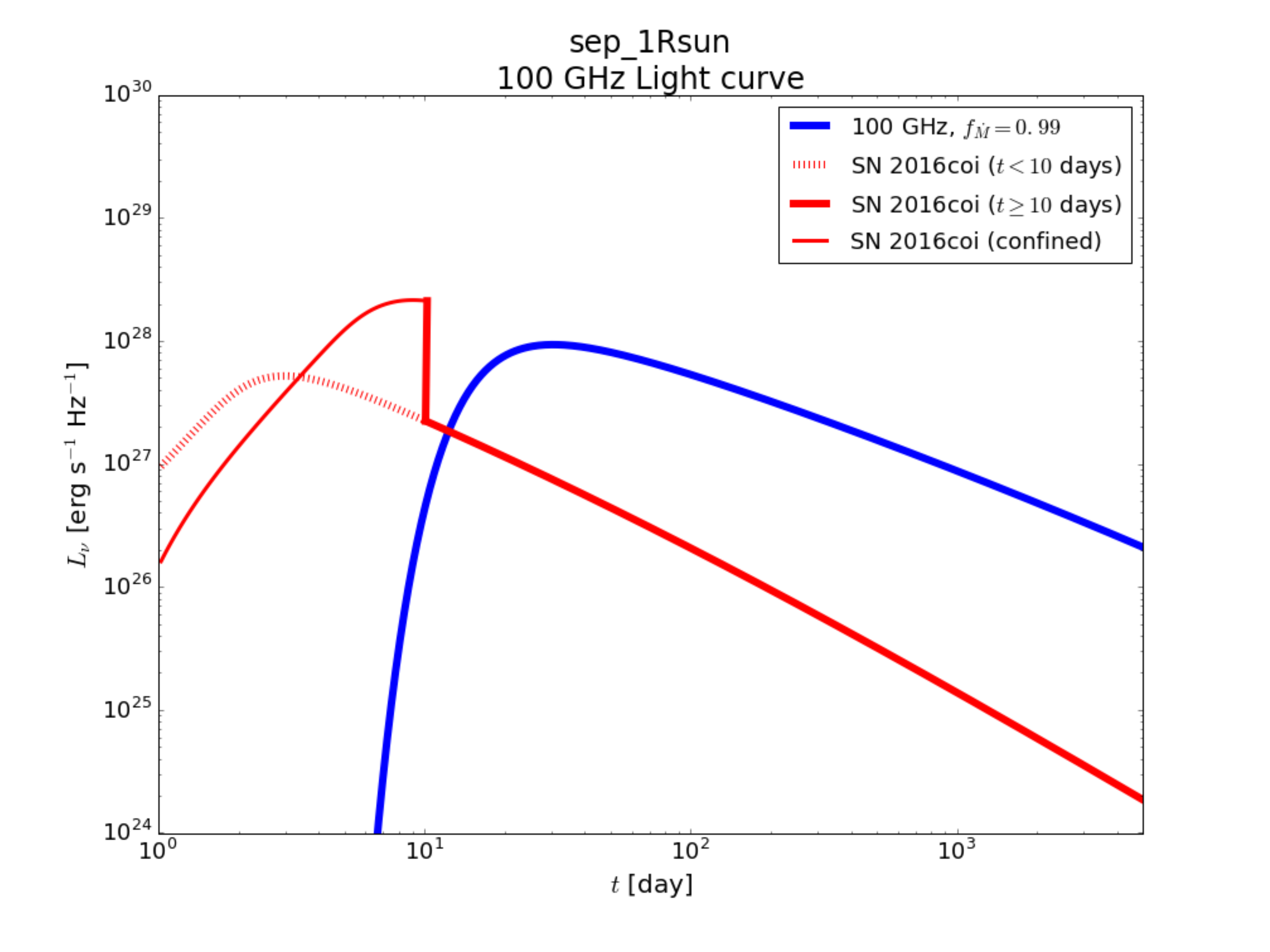}{0.4\textwidth}{(b)}
}
\gridline{
\fig{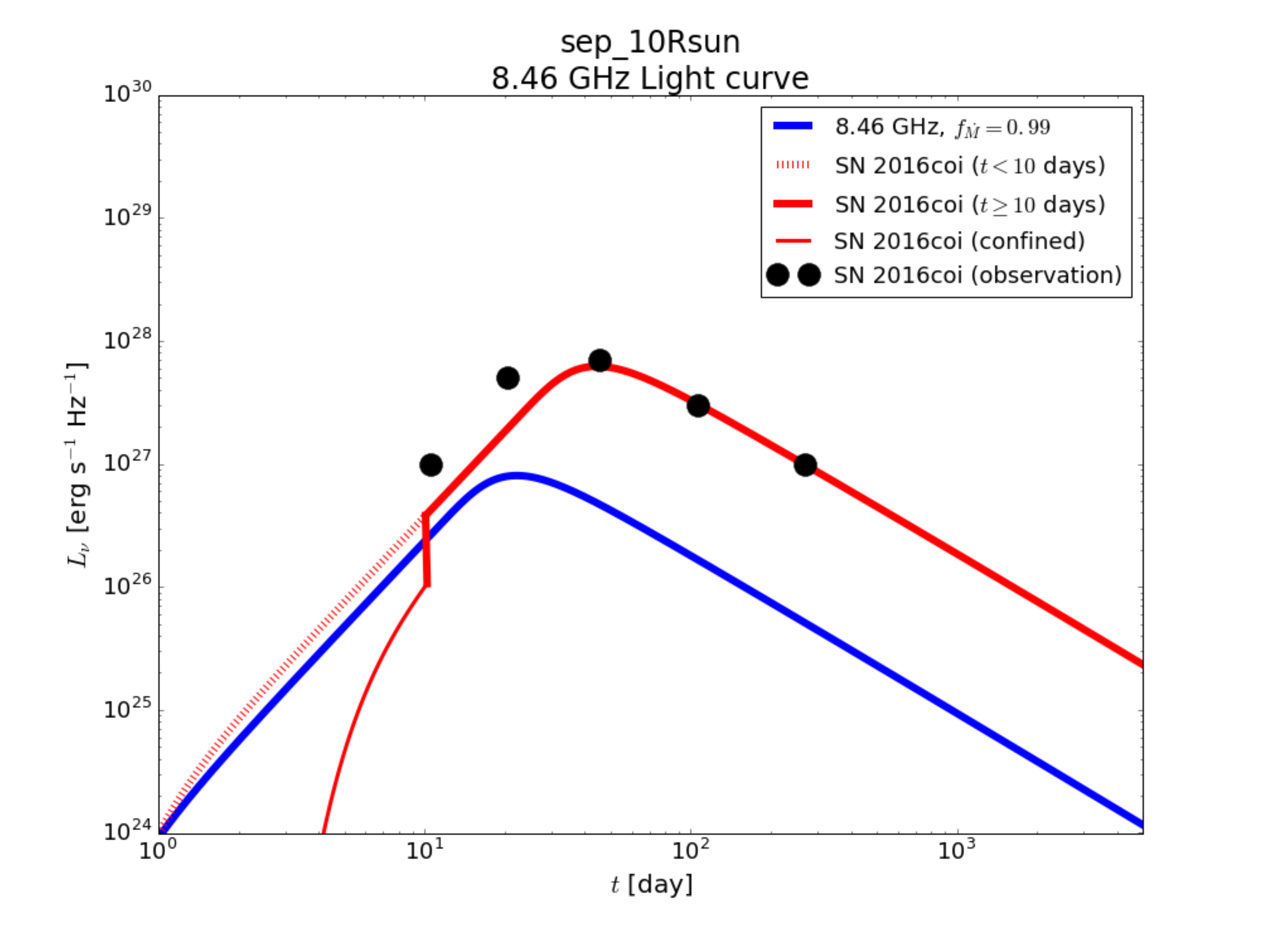}{0.4\textwidth}{(c)}
\fig{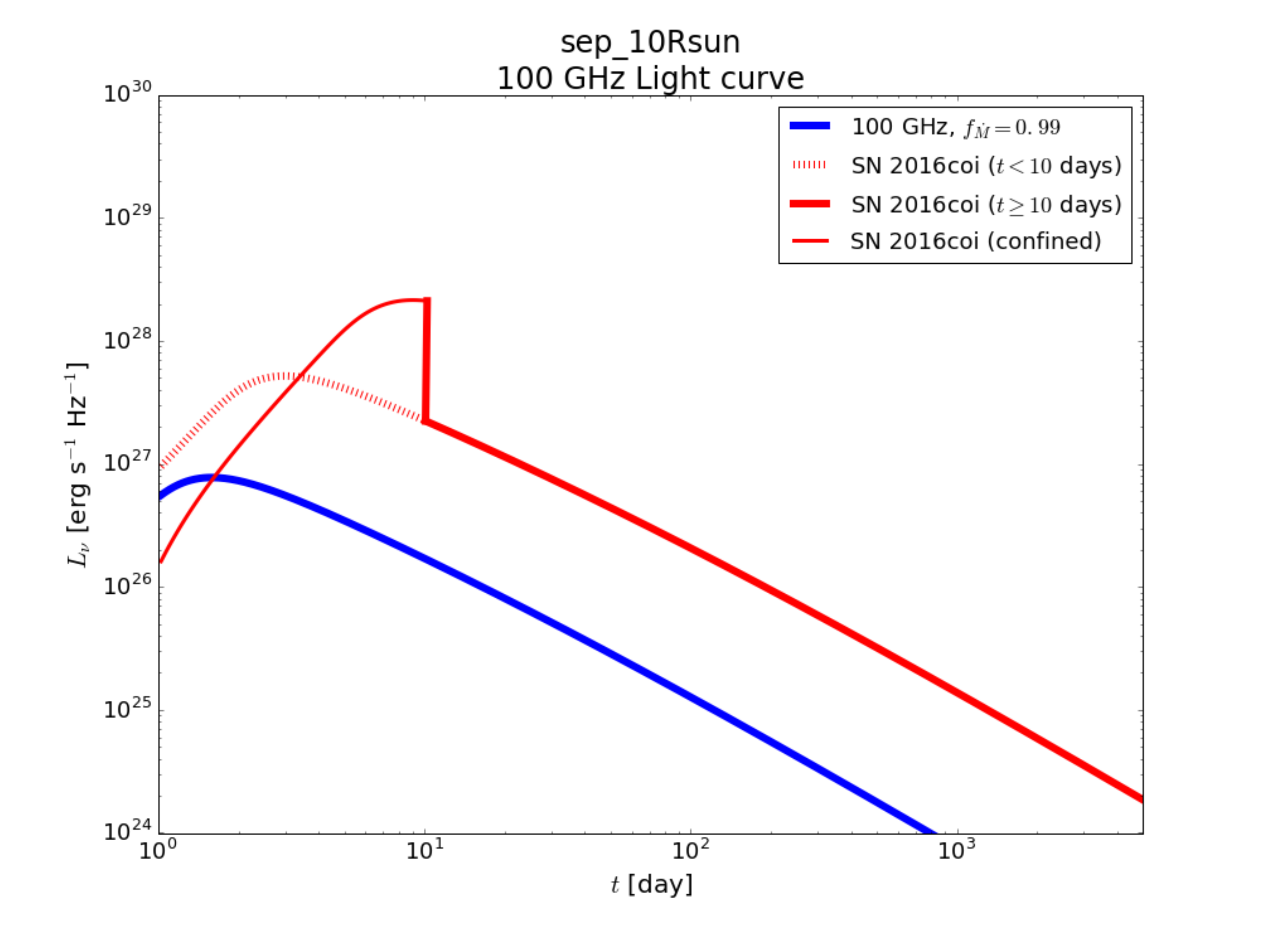}{0.4\textwidth}{(d)}
}
\caption{The modeled radio light curves of SN 2016coi, as compared with our ultra-stripped SN models. The left panels show the radio light curves in the centimeter range (8.46 GHz), while the right panels are models in the millimeter range (100 GHz). The thick red lines show the synthesized radio light curves for SN 2016coi with the mass-loss rate $\dot{M}_{\rm CSM} = 10^{-4} M_\odot{\rm yr}^{-1}$, {for} $t<10$ days (dotted) {or} $t\geq 10$ days (solid). The thin red lines within 10 days indicate the radio light curves {computed with} the putative confined CSM around the SN 2016coi with the mass-loss rate $\dot{M}_{\rm CSM} = 10^{-3} M_\odot{\rm yr}^{-1}$. The black points in the left panels show the observational data of SN 2016coi in the centimeter range \citep{2019ApJ...883..147T}. The ultra-stripped SN models sep\_1Rsun (Top) and sep\_10Rsun (Bottom), with $f_{\dot{M}}=0.99$ as the fiducial value, are plotted with the solid blue lines.}
\label{fig:16coi}
\end{figure*}

Figure \ref{fig:16coi} shows the modeled radio light curves of SN 2016coi, as compared with our fiducial ultra-stripped SN models ($f_{\dot{M}}=0.99$). For the physical parameters of SN 2016coi, we adopt the following values: $M_{\rm ej} = 4.0M_\odot, E_{\rm kin} = 7\times 10^{51} {\rm erg}$, and $\dot{M}_{\rm CSM} = 1.0\times 10^{-4} M_\odot {\rm yr}^{-1}$, as derived by \cite{2019ApJ...883..147T}. Assuming that the progenitor of SN 2016coi would have the confined CSM with the mass-loss rate $\dot{M}_{\rm CSM} = 1.0\times 10^{-3} M_\odot {\rm yr}^{-1}$ up to $\sim 10^{15}$ cm, we also examine the radio emission produced by the collision between the SN ejecta and the confined CSM. This setup corresponds to the situation that the confined CSM will be swept up by the shock within $\sim 10$ days. The observational data for the radio signals from SN 2016coi at 8.46 GHz are also plotted in Figure \ref{fig:16coi}, which {are} roughly consistent with our {model light curve for SN 2016coi}.

The characteristic radio signal produced by the confined CSM appears only within $\sim 10$ days \citep[thin red lines in Figure \ref{fig:16coi}, see also][]{2019ApJ...885...41M}. It is damped due to the FFA in the centimeter range, while it is strongly enhanced in the millimeter range. On the other hand, the final RLO mass transfer in the ultra-stripped SN progenitor binary models continues for $\gtrsim 10$ years, forming a dense CSM up to $\sim 10^{17}$ cm. This extended CSM produces a slowly-decaying radio emission up to $\sim 1000$ days, unlike the signal from the confined CSM. Therefore, a long-term radio monitoring can solve the degeneracy between the RLO {expected} in the ultra-stripped SN progenitor binary {evolution} and the pre-SN {enhanced} mass loss seen in some core-collapse SNe. 

In addition, we note that the radio emissions from SN 2016coi after 10 days are also distinguishable from those from the ultra-stripped SNe. The gap in the mass-loss rate between SN 2016coi and the model sep\_1Rsun results in the {difference in} the radio peak {dates}. For the model sep\_10Rsun, the mass-loss rate is similar to that of SN 2016coi, but the radio luminosity is weaker due to the small explosion energy of the ultra-stripped SN. However, we note that there are some uncertainties on the parameters {describing} the shock acceleration such as $\epsilon_e$ and $\epsilon_B$.
}

\subsection{Event rate and Detectability}
The fraction of ultra-stripped SNe is {suggested} to be 0.1 - 1 percent of the total {number of} SNe \citep{2013ApJ...778L..23T}. If transient observational facilities are able to completely detect all of the SNe within the distance $D_{\rm max}$, the detection rate of ultra-stripped SNe is estimated as $\sim 10 (D_{\rm max}/300$ Mpc$)^3$ yr$^{-1}$ \citep{2019ApJ...882...93H}, corresponding to once per one month. {At} the typical distance of 300 Mpc for the ultra-stripped SNe (corresponding to iPTF 14gqr), the expected flux density of {the radio signal} is $\sim 0.1 (L_{\rm max}/10^{28} {\rm erg}\ {\rm s}^{-1}\ {\rm  Hz}^{-1})$ mJy. {An integration time of  $\sim 5$ minutes is required for Very Large Array (VLA) or Atacama Large Millimeter/submillimeter Array (ALMA) to {detect such a signal with 5$\sigma$ sensitivity}.} We suggest a long-term monitoring of ultra-stripped SN candidates in the centimeter range {with VLA}, or a quick millimeter follow-up observation with {ALMA} as interesting proposals.

\subsection{Model uncertainties}
Before closing this paper, we comment on {a few} uncertainties involved in {our modeling}. First is the velocity of the CSM, which is important for determining the normalization of the CSM density. In this study we have assumed the typical escape velocity of {a} helium star ($u_{\rm} = 10^8$ cm s$^{-1}$). 
However, if the origin of the CSM is dominated by the outflow from the neutron star, the expected signals will become weaker than {those in} the present models.

Second, we employ the free parameter $f_{\dot{M}}$ to describe how much fraction of the {gas transferred} from the helium star will be distributed as the CSM. This parameter could be attributed to two physical processes; one is the {direct} leakage from the Roche Lobe, and the other is the outflow {caused by} the super-Eddington accretion onto the neutron star \citep[see e.g.,][and references therein]{2019A&A...626A..18C}. As discussed in Section \ref{sec:models}, the large {value} is plausible for $f_{\dot{M}}$ from various viewpoints. If the small $f_{\dot{M}}$ {{is}} realized, then the companion NS would experience the gravitational collapse to a black hole, or the accreting gas would stagnate around the NS and the binary would evolve into the common envelope again. The small $f_{\dot{M}}$ thus {neither} produce the ultra-stripped SN, nor form the DNS binary. 

{Finally, we adopt the standard parameters to describe the shock acceleration, $\epsilon_e$ and $\epsilon_B$, to calculate the radio light curves. As noted in Section \ref{sec:B_and_N}, the uncertainty in these parameters will affect the peak date and the optically thin, decaying luminosity of the radio emission. However, we emphasize that the optimized combinations of the observational epoch and frequency for detecting the radio signals from the ultra-stripped SN suggested in this paper would hardly be affected by the uncertainty in these shock acceleration parameters.}

\section{Summary}\label{sec:summary}
A DNS binary imprints the stellar evolution history of massive stars in its formation {process}. Detections of radio pulsars, gravitational waves from a DNS merger and the {associated} kilonova, have uncovered the universal existence of close DNS binaries. The system must experience two SN explosions, and strong binary interaction, in the evolution process. The ultra-stripped SN scenario has been proposed as a promising system to form a DNS binary, thanks to its small ejecta mass. Recent transient observations have discovered some candidates for the ultra-stripped SNe, {and the nature of the candidates has been investigated in detail. However, an ultra-stripped SN progenitor {system} may have large separation, {and then} a remnant DNS binary {would not} merge within the cosmic age. Observational properties {of ultra-stripped SNe in the optical range} are not sensitive to the separation of the remnant DNS binary. An alternative method for investigating the binary separation and the possibility of the remnant DNS merger within the cosmic age is thus required.} 

We have focused on the mass-transfer rate of the ultra-stripped SN progenitor binary, which is {highly sensitive to} the orbital separation. {The high mass-transfer rate in the ultra-stripped SN progenitor binary results from the strong binary stripping associated with the small binary separation, and will be directly linked to the high CSM density around the progenitor}. In such a circumstance, radio emission induced by the {SN-CSM} interaction should be {strong}, and this {will} become a {potential} tracer of the mass-transfer rate and the separation of the progenitor binary.

Guided by the stellar evolution models developed by \citetalias{2015MNRAS.451.2123T}, we have analytically calculated the radio emission from the ultra-stripped SNe. We have shown that the peak luminosities both in the centimeter and millimeter ranges are {high} in {some of} the models with small separations. A strong radio signal can thus {indicate that} the remnant DNS binary can merge within the cosmic age. Furthermore, we have also suggested an optimized combination of the time and frequency windows to study the radio signals from the ultra-stripped {SNe}. {The centimeter emission in the late epoch ($\gtrsim 100$ days) and the millimeter emission in the early epoch ($\lesssim 30$ days) serve as potential {probes} for investigating the nature of the remnant DNS binary}.

\acknowledgments

The authors thank Takashi J. Moriya and Yudai Suwa for providing us with the data of the progenitor of ultra-stripped SNe. K.M. acknowledges support by JSPS KAKENHI Grant ({20H00174, 20H04737}, 18H04585, 18H05223, and 17H02864).

\bibliography{manuscript}

\end{document}